\documentclass[journal,11pt,onecolumn]{IEEEtran}

\usepackage{algorithmicx}
\usepackage{algpseudocode}
\usepackage{algorithm, setspace}
\algdef{SE}[DOWHILE]{Do}{doWhile}{\algorithmicdo}[1]{\algorithmicwhile\ #1}%
\usepackage{amsmath,graphicx}
\usepackage{dsfont}
\usepackage{amsfonts}
\usepackage[utf8]{inputenc}
\usepackage[usenames, dvipsnames]{color}
\usepackage{subfigure}
\usepackage{pifont}
\usepackage{multirow}
\newcommand{\cmark}{\ding{51}}%
\newcommand{\xmark}{\ding{55}}%

\usepackage[noadjust]{cite}
\usepackage{hyperref}
\usepackage{array}
\newcolumntype{L}[1]{>{\raggedright\let\newline\\\arraybackslash\hspace{0pt}}m{#1}}
\newcolumntype{C}[1]{>{\centering\let\newline\\\arraybackslash\hspace{0pt}}m{#1}}
\newcolumntype{R}[1]{>{\raggedleft\let\newline\\\arraybackslash\hspace{0pt}}m{#1}}
\begin{document}
\title{Ship Wake Detection in SAR Images\\ via Sparse Regularisation}
\author{Oktay~Karakuş,
Igor~Rizaev,
        Alin~Achim,
        \thanks{This work was supported by the Engineering and Physical Sciences Research Council (EPSRC) under grant EP/R009260/1 (AssenSAR).}
        \thanks{Oktay Karakuş , Igor Rizaev
        and Alin Achim are with the Visual Information Laboratory, University of Bristol, Bristol BS1 5DD, U.K. (e-mail: o.karakus@bristol.ac.uk; i.g.rizaev@bristol.ac.uk: alin.achim@britol.ac.uk)}
}

\maketitle
\begin{abstract}
In order to analyse synthetic aperture radar (SAR) images of the sea surface, ship wake detection is essential for extracting information on the wake generating vessels. One possibility is to assume a linear model for wakes, in which case detection approaches are based on transforms such as Radon and Hough. These express the bright (dark) lines as peak (trough) points in the transform domain. In this paper, ship wake detection is posed as an inverse problem, which the associated cost function including a sparsity enforcing penalty, i.e. the \textit{generalized minimax concave} (GMC) function. Despite being a non-convex regularizer, the GMC penalty enforces the overall cost function to be convex. The proposed solution is based on a Bayesian formulation, whereby the point estimates are recovered using \textit{maximum a posteriori} (MAP) estimation. To quantify the performance of the proposed method, various types of SAR images are used, corresponding to TerraSAR-X, COSMO-SkyMed, Sentinel-1, and ALOS2. The performance of various priors in solving the proposed inverse problem is first studied by investigating the GMC along with the $L_1$, $L_p$, nuclear and total variation (TV) norms. We show that the GMC achieves the best results and we subsequently study the merits of the corresponding method in comparison to two state-of-the-art approaches for ship wake detection. The results show that our proposed technique offers the best performance by achieving 80\% success rate.
\end{abstract}
\IEEEpeerreviewmaketitle
\section{Introduction}\label{sec:intro}
Accurate characterisation of sea surface condition is not only important in isolation, but also in the detection and characterisation of ship wakes. These provide key information for tracking (illegal) vessels and are also useful in classifying the characteristics of the wake generating vessel. Until recently, one of the main factors hampering research into sea surface modelling was the lack of data of sufficiently high resolution (pixels need to be typically smaller than few meters) and accuracy. Synthetic aperture radar (SAR) technologies have however shown remarkable progress in recent years and the availability of remotely sensed data of the Earth and sea surface is continuously growing. Several European missions (e.g., the Italian COSMO/SkyMed, the German TerraSAR-X, or the British NovaSAR mission) have developed a new generation of satellites exploiting synthetic aperture radar (SAR) to provide spatial resolutions previously unavailable from space-borne remote sensing.

In all SAR images, bright areas correspond to high radar cross-section per unit area and dark areas to low radar cross-section. When imaging sea surface, high returns are caused by enhanced surface roughness or large wave steepness, either via specular reflection or Bragg scattering. Whereas specular reflections are usually caused by breaking waves, the Bragg scatterers are short gravity waves or capillary waves depending on the wavelength of the SAR sensor itself. Sea waves become visible in SAR images due to the Bragg scattering of SAR electromagnetic waves by small scale capillary and gravity-capillary waves that propagate on the surface of larger waves \cite{alpers1979effect}. These include swells, coherent Kelvin waves, random sea waves as well as their mixture \cite{reed1990hydrodynamics,reed2002ship}. SAR images of moving ships exhibit some characteristic patterns which are directly determined by different ship wake formations. These are typically considered to fall in one of three categories: (i) turbulent wakes, (ii) surface waves created by ships and (iii) ship-generated internal waves. Ship-generated surface waves can in turn be split into two sub-categories \cite{lyden1988synthetic, wright1968new}, the first being the short (centimeter scale) waves, while the second includes the (decameter scale) waves forming the classical Kelvin wake system \cite{wright1974microwave}. The former are observed in SAR images through the Bragg scattering mechanism and appear as bright, narrow V-wakes due to the resonant interaction of the transmitted radar waves and ocean surface waves \cite{wright1966backscattering}. The two-scale composite model has been employed to simulate SAR images of rough sea surfaces with embedded Kelvin wake structures in \cite{zilman2015detectability}, while Fujimura et al. performed a validation study for simulated ship wakes via computational dynamics in \cite{fujimura2010numerical}.

Since ship wakes can be modelled as linear structures, corresponding detection methods are mostly based on linear feature extraction approaches, such as the Hough or Radon transforms, both of which create bright peaks in the transform domain for bright lines, and troughs for dark lines. Due to its high computational cost, the Hough transform has attracted less interest than the Radon transform \cite{graziano1}. Thanks to the lower computational complexity of its inverse, the Radon transform is widespread in ship wake detection applications and has been first utilised by Murphy \cite{murphy1986linear}. The Radon transform has however a couple of drawbacks, e.g. bright pixels belonging to ships may cause false detections \cite{grazianoex}. To address this, enhancing the Radon domain information is common practice in the literature. Rey et al. \cite{rey1990} have proposed a method which combines the Radon transform and Wiener filtering to increase the detectability of the peaks in Radon domain. Tunaley \cite{tunaley2003estimation} used a method which restricts the search area in Radon domain. Eldhuset \cite{eldhuset1996automatic} proposed an automatic ship and wake detection method whereby the detection performance is characterised by the number of lost and false wakes (wake-like features). Based on ship beam and speed estimation, Zilman et al. \cite{zilman2004speed} have applied an enhancement operation to the Radon transform of the observed (noisy) image. Courmontagne \cite{courmontagne2005improvement} used a method based on a combination of Radon transform and stochastic matched filtering for wake detection. Kuo and Chen \cite{kuo2003application} have proposed a ship wake detection method using wavelet correlators. Zilman et al. \cite{zilman2015detectability} proposed a SAR image simulator, including ship wakes, and studied the performance of their ship wake detection method previously published in \cite{zilman2004speed}. Recent studies based on $L_2$ regularised logistic regression \cite{tings2018comparison} and low-rank plus sparse decomposition \cite{biondi2018low} also addressed the ship wake detection problem in SAR images. Graziano et. al. \cite{grazianoradonMerit,grazianoex,graziano1,graziano2} have proposed wake detection methods which deal directly with the noisy image without performing any preliminary enhancement. The authors have first created ship-centred-masked image tiles and performed a restricted area search in their Radon representation. The restricted area is the area that lies between two sine waves, the peak points of which have been selected using real ancillary data, local map, ship clusters, and the local traffic statistics.

The inverse problem formulation for line detection has been first proposed by Aggarwal and Karl \cite{aggarwal2006line}. The main advantage of formulating line detection as an inverse problem is the subsequent use of a regularisation framework, which allows the incorporation of prior information about the object of interest. Anantrasirichai et al. \cite{anantrasirichai2017line} have further investigated the inverse problem formulation for B-line detection in lung ultrasound images. Although the inverse problem solution creates an enhanced image, this step is different from operations like despeckling, which require statistical assumptions \cite{achim2006sar, moser2006sar} or statistical model selection \cite{karakucs2019generalized} and is well studied in the literature \cite{pivzurica1999image, pizurica2001despeckling, achim2003sar, datcu2007wavelet, molina2010gibbs}.

In this paper, we propose a novel approach to ship wake detection in SAR images which is based on an inverse problem formulation. The main contribution of this paper is to propose an innovative approach based on sparse regularisation to obtain the Radon transform of the image, in which the linear features are enhanced. The solution to the inverse problem involves Bayesian methodology which leads to a \textit{maximum a-posteriori} (MAP) estimator. Our proposed cost function uses the \textit{generalised minimax concave} (GMC) penalty of Selesnick \cite{selesnick2} as regularisation term and investigates its merit in comparison to the \textit{total variation} (TV), \textit{nuclear}, $L_1$ and $L_p$ norms. The use of the GMC penalty demonstrates the advantages of non-convex sparse regularisation while allowing the cost function to remain convex. We use SAR images of the sea surface from different sources, including TerraSAR-X, COSMO-SkyMed, SENTINEL-1 and ALOS2 to test the performance of the proposed method. For the ship wake detection step, we use a method which performs detection in the Radon domain as proposed in \cite{graziano1,graziano2}. In this study, contrary to \cite{graziano1,graziano2} the proposed method detects ship wakes, independently of real information about the ships and the environment, by selecting the required parameters directly from the observed SAR images. MAP estimates for the images are obtained using the \textit{forward-backward }(FB) method and \textit{the two-step iterative shrinkage/thresholding} (TwIST) algorithm \cite{bioucas2007new}.

The rest of the paper is organised as follows: the image formation model for ship wakes identification, the MAP estimation, GMC regularisation and priors employed are discussed in Section \ref{sec:Methods}. The detection algorithms and SAR data sets are presented in Sections \ref{sec:Detection} and \ref{sec:Data}, respectively. Experimental studies and results are provided in Section \ref{sec:Exp}. Section \ref{sec:conc} concludes the paper with a brief summary.

\section{Theoretical Preliminaries}\label{sec:Methods}
\subsection{Ship Wakes and Image Formation Model}\label{sec:sec2}

As discussed briefly in the previous section, a moving ship in deep sea typically creates three different types of wakes. The central dark streak is called \textit{turbulent wake}. Generally, this central dark streak is surrounded by two bright arms called narrow V-wake, which lies either side of the turbulent wake within the half angle from $1.5^{\circ}$ up to $4^{\circ}$ \cite{zilman2004speed, graziano1}. Lastly, two outer arms are known as Kelvin wake and limit the signatures of the moving ship on each side of the turbulent wake with a maximum half angle of $19.5^{\circ}$. The Kelvin wake half angle can be somewhat smaller in real SAR images, e.g. between $10^{\circ}$ and $19.5^{\circ}$ \cite{reed1990hydrodynamics,zilman2015detectability}.

\begin{figure}[ht!]
\centering
\includegraphics[width=.75\linewidth]{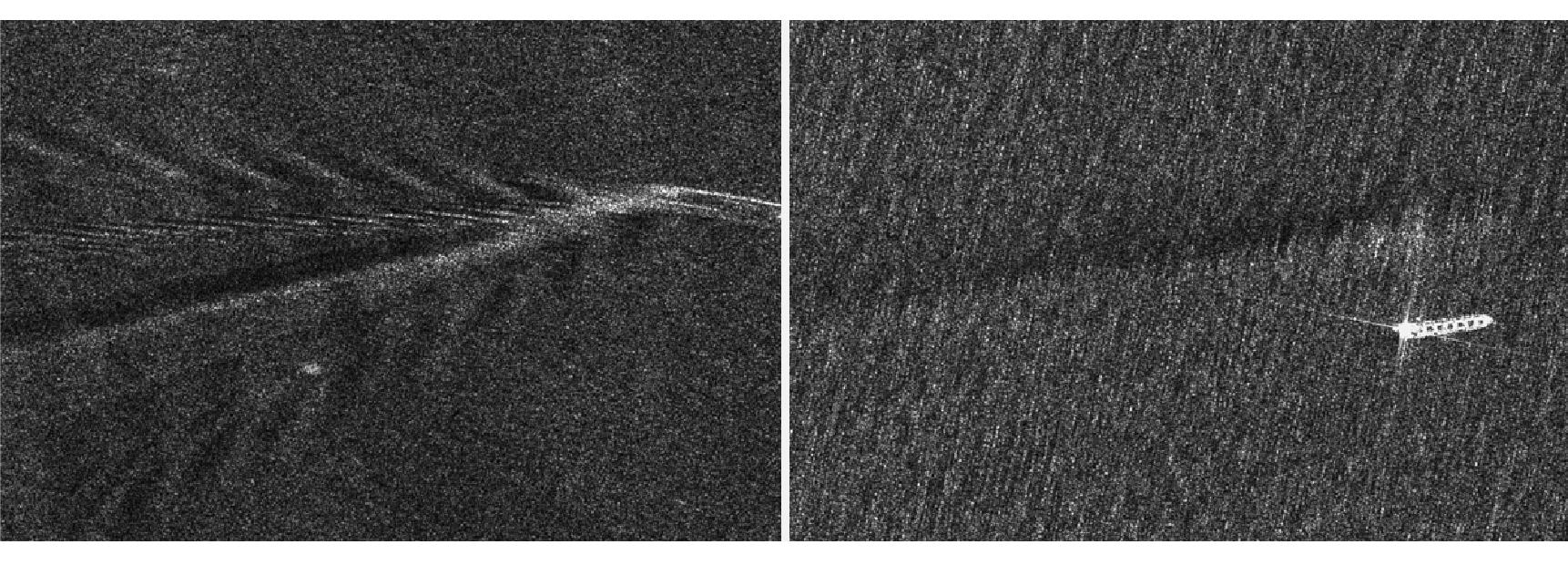}\\
\includegraphics[width=.75\linewidth]{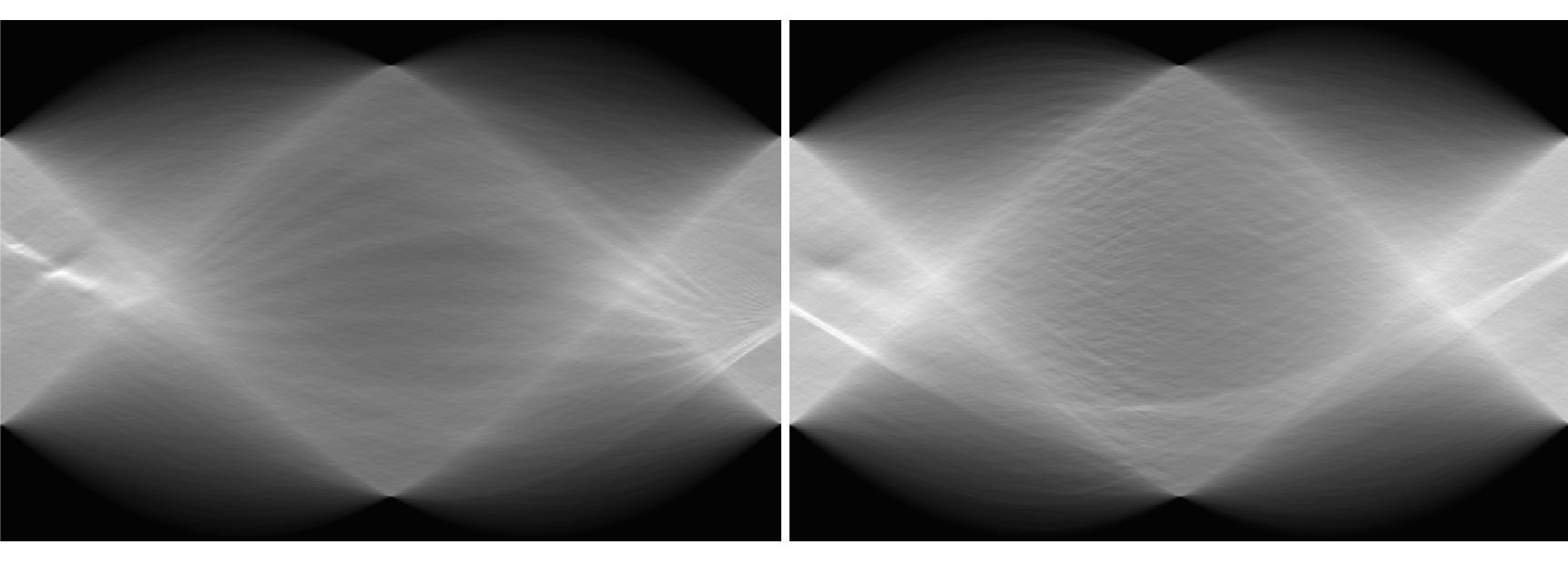}
\caption{Example images including ship wakes and their Radon transform. (Upper Left) - Image includes all 5 ship wakes. (Upper Right) - Image does not include Kelvin wake. (Lower row) - Radon transforms of the images above. Both images are TerraSAR-X Stripmap products with HH polarisation and 3m$\times$3m resolution.}
\label{fig:ExSAR2}
\end{figure}

For the purpose of this study, we consider each arm of narrow-V and Kelvin wakes as a separate wake and refer to them as the first and second wakes in the order of their detection. Consequently, we attempt to detect \textit{five} linear wake-corresponding structures, which are the turbulent wake (1), the first (2) and second (3) narrow V - and the first (4) and second (5) Kelvin - wakes. In most cases, not all five wakes are however visible in SAR images as is generally the case with one of the narrow V and/or Kelvin wakes. In Figure \ref{fig:ExSAR2}, two example SAR images are depicted. In particular, in the upper left image, all five wakes are visible, whereas the one on the upper right does not have visible Kelvin wake.

Since we model ship wakes as linear features, the SAR image formation can be expressed in terms of its Radon transform as
\begin{align}\label{equ:invprob}
  Y = \mathcal{C}X + W,
\end{align}
where $Y$ is the $M\times M$ SAR image, $X(r, \theta)$ is the image in Radon domain, $W$ refers to the additive noise and the operator $\mathcal{C} = \mathcal{R}^{-1}$ is the inverse Radon transform. $X(r, \theta)$ represents lines as a distance $r$ from the centre of $Y$ and an orientation $\theta$ from the horizontal axis of $Y$.

In image processing applications, to compute an integral of the intensities of image $Y(i, j)$ over the hyperplane which is perpendicular to $\theta$ leads to the Radon transform $X(r, \theta)$ of the given image $Y$. It can also be defined as a projection of the image along the angles, $\theta$. Hence, for a given image $Y$, the general form of the Radon transform ($X=\mathcal{R}Y=\mathcal{C}^TY$) is
\begin{align}\label{equ:R}
  X(r, \theta) = \int_{\mathds{R}^2} Y(i, j) \delta(r - i\cos\theta - j\sin\theta) didj
\end{align}
where $\delta(\cdot)$ is the Dirac-delta function.

The inverse Radon transform ($Y=\mathcal{C}X$) of the projected image $X$ can be obtained from the filtered back-projection \cite{avinash1988principles} algorithm as
\begin{align}\label{equ:C}
  g(r, \theta) &= \mathcal{F}^{-1} \left[ |v| \mathcal{F} \left[ X(r, \theta) \right] \right]\\
  Y(i, j) &= \int_{0}^{\pi} \int_{\mathds{R}} g(r, \theta) \delta(r - i\cos\theta - j\sin\theta) drd\theta
\end{align}
where $v$ is the radius in Fourier transform, and $\mathcal{F}\left[\cdot\right]$ and $\mathcal{F}^{-1}\left[\cdot\right]$ refer to forward and inverse Fourier transforms, respectively. In this paper, we use discrete operators $\mathcal{R}$ and $\mathcal{C}$ as described in \cite{kelley1993fast}.

\subsection{Bayesian Inference in Inverse Problems}\label{sec:Bayes}
Assume the aim is to extract information about the unknown signal $x$ given an observation signal $y$. Then, suppose a statistical model which defines the relation of $y$ to $x$ can be expressed. This statistical model is referred to as the likelihood $p(y|x)$. Obtaining $x$ from $y$ might contain numerous uncertainties and thus extracting information belonging to $x$ would be ill-posed and problematic. The Bayesian inference framework helps to reduce these uncertainties by employing knowledge on $x$, namely the prior of $x$, which can promote structural properties such as sparsity.

Hence, incorporating this prior in conjunction with the observed statistical model, creates the knowledge on $x$ given $y$, namely the posterior distribution $p(x|y)$ via Bayes' theorem:
\begin{align}\label{equ:bayes1}
  p(x|y) = \dfrac{p(y|x)p(x)}{\int p(y|x)p(x)dx}
\end{align}
where, the denominator $\int p(y|x)p(x)dx$ is the marginal likelihood $p(y)$ which is not related to $x$ and assumed to be constant according to $x$. Since the aim is to extract $x$, we can write the unnormalized posterior distribution as
\begin{align}\label{equ:bayes11}
  p(x|y) \propto p(y|x) p(x).
\end{align}

Generally, posterior distributions are assumed to be log-concave as
\begin{align}\label{equ:bayes2}
  p(x|y) \propto \exp{\{-F(x)\}}
\end{align}
where $F(x)$ is a convex function. Estimating $x$ directly using the posterior density can be intractable especially for high dimensional cases. Hence, the \textit{maximum a-posteriori} (MAP) estimator maximises the posterior $p(x|y)$ to obtain a point estimate $\hat{x}$. The MAP estimator for the unknown $x$ is
\begin{align}\label{equ:bayes3}
  \hat{x}_{\text{MAP}} = \arg\max_{x} p(x|y) = \arg\min_{x} F(x).
\end{align}

In cases when $F(x)$ is not convex, i.e. the posterior will no longer be log-concave, the minimizer will not be convex but nevertheless can be solved via proximal operators \cite{green2015bayesian}.

According to the SAR image formation defined in (\ref{equ:invprob}), the posterior distribution of the desired line image $X$ with respect to the observed noisy SAR image $Y$ can be written using (\ref{equ:bayes11}) as
\begin{align}\label{equ:post}
  p(X|Y) \propto p(Y|X)p(X).
\end{align}

Assuming an i.i.d. standard normal noise case, the likelihood distribution $p(Y|X)$ can be expressed as
\begin{align}\label{equ:lhd}
  p(Y|X) \propto \exp \{ -\|Y - \mathcal{C}X\|_2^2 \}.
\end{align}

In this paper, we assume priors to be of exponential form as, $p(X) \propto \exp \{-\lambda\psi(X)\}$. Thus, the cost function in (\ref{equ:bayes3}) becomes
\begin{align}\label{equ:F}
  F(X) \propto \|Y - \mathcal{C}X\|_2^2 + \lambda\psi(X),
\end{align}
where $\lambda$ is the scale parameter, namely the regularisation constant.

\subsection{Generalised Minimax Concave (GMC) Penalty}\label{sec:GMC}
The GMC regularisation has been proposed by Selesnick \cite{selesnick2} as a sparsity enforcing penalty in inverse problems. It exploits the advantages of using non-convex penalties as well as preserving the convexity of the cost function. It is based on $L_1$ norm and the generalised Huber function.

In particular, the minimax concave (MC) penalty in the univariate case is
\begin{align}\label{equ:MC1}
  \psi(t) =
  \begin{cases}
      |t| - \dfrac{1}{2}t^2 & |t|\leq 1 \\
      \dfrac{1}{2} & |t|\geq 1.
   \end{cases}
\end{align}

The relation between the MC penalty and the Huber function $s(t)$ can be written as
\begin{align}\label{equ:MC2}
  \psi(t) = |t| - s(t).
\end{align}

For a scalar, $b\neq0$, the scaled MC penalty, $\psi_b(t)$ can be written as scaled Huber function, $s_b(t)$ as
\begin{align}\label{equ:MC3}
  \psi_b(t) = |t| - s_b(t).
\end{align}

All these definitions are based on the univariate case, but they can be adapted to the multivariate case. If we assume a scaling matrix, $B$, then the generalised Huber function, $S_B(t)$ can be defined as \cite{selesnick2}
\begin{align}\label{equ:GenHuber}
  S_B(t) = \inf_v \{ \|v\|_1 + \frac{1}{2}\|B(t - v)\|_2^2 \}.
\end{align}

Combining (\ref{equ:MC3}) and (\ref{equ:GenHuber}) gives the generalised MC (GMC) penalty function as
\begin{align}\label{equ:MC4}
  \psi_B(t) = \|t\|_1 - S_B(t).
\end{align}

The scaling matrix $B$ should be selected in relation to the inverse Radon operator $\mathcal{C}$ to provide the convexity of the cost function as
\begin{align}\label{equ:MC5}
  B = \sqrt{\dfrac{\gamma}{\lambda_1}} \mathcal{C}
\end{align}
where $\lambda_1$ is the scale parameter of the GMC prior and $\gamma$ is a parameter which controls the non-convexity. Note that for $0\leq\gamma\leq1$, $B$ ensures the convexity of the cost function. The nominal range of $0.5\leq\gamma\leq0.9$ should be used for better performance \cite{selesnick2}.

Thus, the GMC sparse prior can be written as
\begin{align}
  p_1(X) &\propto \exp \{-\lambda_1\psi_B(X)\}\\
  &\propto\exp\{ -\lambda_1\left(\|X\|_1 - S_B(X) \right) \}.\label{equ:GMCprior}
\end{align}

As GMC regularisation does not have an explicit formulation, the minimisation problem with the cost function in (\ref{equ:F}) can be solved using proximal operators. Thus, we rewrite the cost function as
\begin{equation}
\begin{aligned}\label{equ:MAPGMC2}
  F(X, v) = \|Y - \mathcal{C}X\|_2^2 & + \lambda_1 \|X\|_1 - \lambda_1 \|v\|_1 - \gamma \|\mathcal{C}(X - v)\|_2^2.
\end{aligned}
\end{equation}
which leads to a minimax optimisation problem
\begin{align}\label{equ:MAPGMC}
  \hat{X}_{\text{MAP-GMC}} = \arg \min_X \max_v F(X, v).
\end{align}

The solution to this problem can be obtained using the forward-backward (FB) algorithm \cite{selesnick2}. The FB algorithm for GMC regularised cost functions requires only a couple of computational steps and soft-thresholding, which is the proximal operator for $L_1$ based regularisation. The FB algorithm to solve (\ref{equ:MAPGMC}) is given in Algorithm \ref{alg:FB}. For the purpose of this study, we chose the maximum number of iterations $MaxIter = 1000$, which was experimentally set. The algorithm stops when the error term reaches $10^{-3}$. This error term $\epsilon^{(i)}$ is calculated for iteration $i$ as
\begin{align}\label{equ:Stop}
   \epsilon^{(i)} = \dfrac{\|X^{(i)} - X^{(i-1)}\|}{\|X^{(i-1)}\|}.
 \end{align}

\begin{algorithm}[h!]
\setstretch{1.1}
\caption{Forward-backward algorithm for GMC regularised cost function }\label{alg:FB}
\begin{algorithmic}[1]
\State \textbf{Input:} $\text{Ship-centered-masked SAR image }Y$
\State \textbf{Input:} $\lambda>0$
\State \textbf{Input:} $0.5\leq\gamma\leq0.9$
\State \textbf{Output:} $\text{Radon image }X$
\State \textbf{Set:} $0<\mu<1.9 \text{ and } i=0$
  \Do
    \State $w^{(i)} = X^{(i)} - \mu\mathcal{C}^T(\mathcal{C}(X^{(i)} + \gamma(v^{(i)} - X^{(i)})) - Y)$
    \State $u^{(i)} = v^{(i)} - \mu\gamma\mathcal{C}^T\mathcal{C}(v^{(i)} - X^{(i)})$
    \State $X^{(i)} = soft(w^{(i)}, \mu\lambda)$
    \State $v^{(i)} = soft(u^{(i)}, \mu\lambda)$
    \State $i++$
  \doWhile{$\epsilon^{(i)} > 10^{-3} \text{ or } i<MaxIter$}
\end{algorithmic}
\end{algorithm}

\subsection{Alternative Sparse Priors}
In this paper, we also investigate other types of priors, which are known to be sparse and common in optimisation problems. The first of which is Laplace (or $L_1$ norm) prior, which is log-concave and given by
\begin{align}\label{equ:L1prior}
  p_2(X) &\propto \exp \{-\lambda_2\psi_2(X)\}\\
  &\propto \exp\{ -\lambda_2\|X\|_1 \}.\label{equ:L1prior2}
\end{align}

We further investigate the (non-convex) $L_p$ norm based prior, which is
\begin{align}\label{equ:Lpprior}
  p_3(X) &\propto \exp \{-\lambda_3\psi_3(X)\}\\
  &\propto\exp\{ -\lambda_3\|X\|_p^p \}\label{equ:Lpprior2}
\end{align}
where $0 < p < 1$.

Another important type of prior is the TV norm, $TV(\cdot)$, which can be defined as \cite{pereyra2016proximal}
\begin{align}\label{equ:TVprior2}
  TV(X) = \| \nabla X \|_1.
\end{align}
where $\nabla$ is the 2-D discrete gradient operator. Hence, the TV norm based prior is given by
\begin{align}\label{equ:TVprior}
  p_4(X) &\propto \exp \{-\lambda_4\psi_4(X)\}\\
  &\propto \exp\{ -\lambda_4 TV(X) \}.\label{equ:TVprior3}
\end{align}

The nuclear norm is an important type of prior for data with low-rank and low singular values. In particular, $\|X\|_*$ is the nuclear norm of $X$ and is defined as the sum of its singular values \cite{pereyra2016proximal}. Here, we define the nuclear norm prior as:
\begin{align}\label{equ:Nucprior}
  p_5(X) &\propto \exp \{-\lambda_5\psi_5(X)\}\\
  &\propto \exp\{ -\lambda_5\|X\|_* \}.\label{equ:Nucprior2}
\end{align}

Replacing $\psi(X)$ in (\ref{equ:F}) with the previously defined priors determines different MAP estimators as
\begin{align}\label{equ:MAPAll}
  \hat{X}_{\text{MAP}-\psi_k} = \arg\min_X \{ \|Y - \mathcal{C}X\|_2^2 & + \lambda_k \psi_k(X) \}
\end{align}
where $k = 2, 3, 4, 5.$

Minimisations of (\ref{equ:MAPAll}) are subsequently carried out with the two-step iterative shrinkage/thresholding (TwIST) algorithm \cite{bioucas2007new}. The TwIST algorithm is given in Algorithm \ref{alg:FB2} where $MaxIter$ and the stopping criterion $\epsilon$ are the same as in the GMC case.
\begin{algorithm}[h!]
\setstretch{1.1}
\caption{TwIST algorithm}\label{alg:FB2}
\begin{algorithmic}[1]
\State \textbf{Input:} $\text{Ship-centered-masked SAR image }Y$
\State \textbf{Output:} $\text{Radon image }X$
\State \textbf{Set:} $\lambda_k>0 \text{ and } \alpha=1.96 \text{ (as in \cite{bioucas2007new})}$
\State \textbf{Set:} $X^{(1)}=\Gamma_{\lambda_k}(X^{(0)}) \text{ and } i = 1$
  \Do
    \State $w^{(i)} = X^{(i)} + \mathcal{C}^T\left(Y - \mathcal{C}X^{(i)} \right)$
    \State $\text{Obtain } \Gamma_{\lambda_k}(w^{(i)}) \text{ for } \psi_k$
    \State $X^{(i+1)} = (1-\alpha)X^{(i-1)} - \alpha X^{(i)} + 2\alpha\Gamma_{\lambda_k}(w^{(i)})$
    \State $i++$
  \doWhile{$\epsilon^{(i)} > 10^{-3} \text{ or } i<MaxIter$}
\end{algorithmic}
\end{algorithm}

The operator $\Gamma_{\lambda_k}(\cdot)$ in Algorithm \ref{alg:FB2} is a shrinkage/thresholding/denoising function. In this paper, we used shrinkage/thresholding operators in minimizers with $L_1$ and $L_p$ for $\Gamma_{\lambda_k}(\cdot)$ as also used in \cite{anantrasirichai2017line}
\begin{align}\label{equ:proxTwIST}
  \Gamma_{\lambda_k}(u) = \frac{|prox_{\psi_k}^{\lambda_k}(u)|}{|prox_{\psi_k}^{\lambda_k}(u)| + \lambda_k}u \quad \text{for } k = 2 \text{ or } 3,
\end{align}
and the denoising operators for the TV and nuclear norm priors,
\begin{align}\label{equ:proxTwIST2}
  \Gamma_{\lambda_k}(u) = prox_{\psi_k}^{\lambda_k}(u) \quad \text{for } k = 4 \text{ or } 5.
\end{align}
where $prox_{\psi_k}^{\lambda_k}(\cdot)$ is the Moreau proximal operator for $\psi_k(\cdot)$.

The soft thresholding operation is the proximal operator for $L_1$ norm, whereas for $L_p$ norm the proximal operator is computed with an iterative algorithm called generalised soft thresholding (GST) \cite{zuo2013generalized,anantrasirichai2017line}. The proximal operator for nuclear norm is obtained via singular value soft thresholding as in \cite{pereyra2016proximal} and for TV norm it is efficiently computed by using Chambolle's method in \cite{chambolle2004algorithm}.

\section{Ship Wake Detection}\label{sec:Detection}
\begin{figure}[ht]
\centering
\includegraphics[width=.7\linewidth]{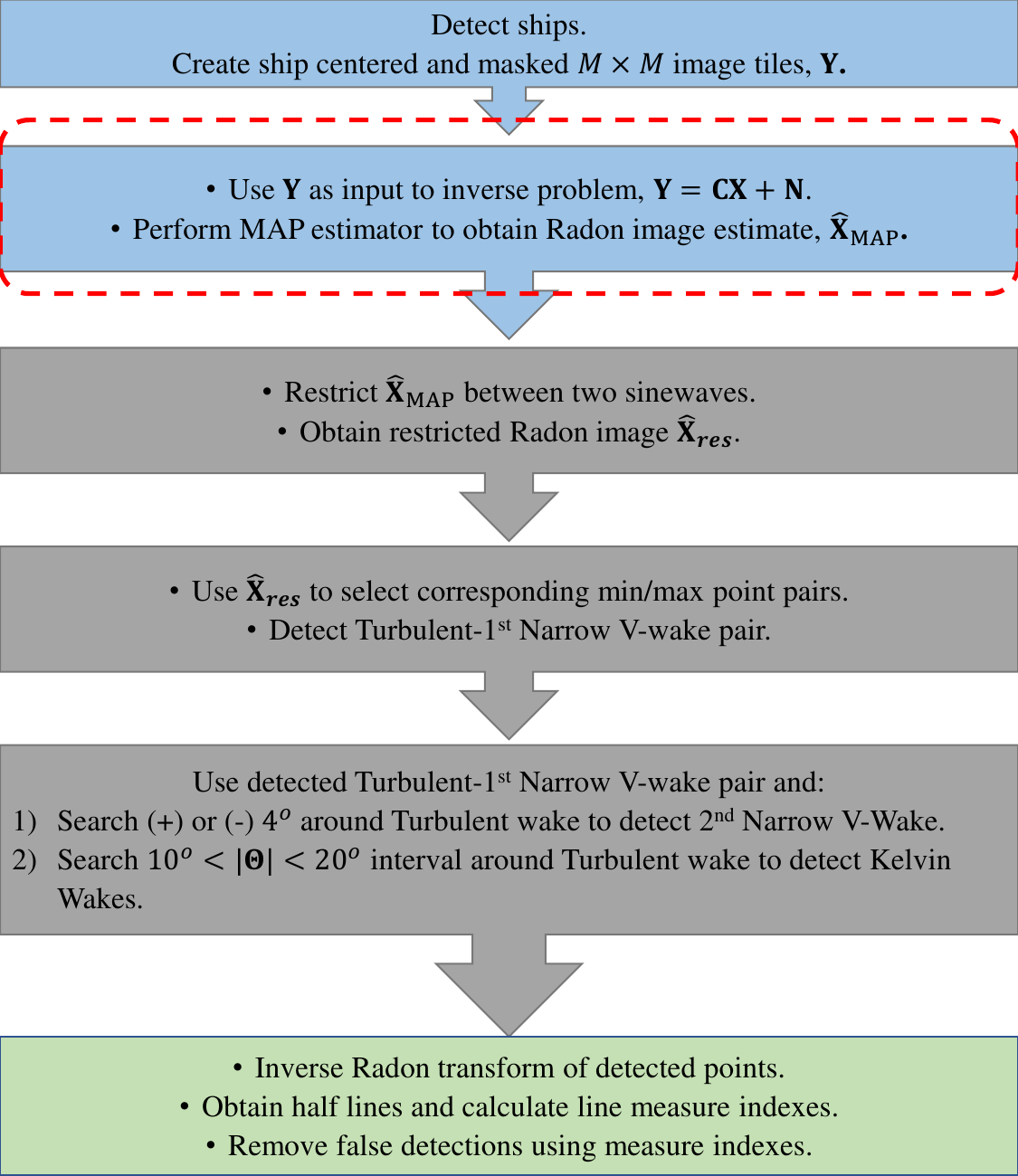}
\caption{Block diagram of the proposed method.}
\label{fig:flow}
\end{figure}

Our detection algorithm includes three important steps as shown with different coloured rectangles in Figure \ref{fig:flow}: 1) Pre-processing and inverse problem solution (blue rectangles), 2) detection of wakes in the Radon domain (grey rectangles) and 3) validation in the spatial domain (green rectangle).The blue rectangle within the red dashed-lined shape in Figure \ref{fig:flow} constitutes the main contribution of this paper, while the detection/validation steps are inspired by the method in \cite{graziano1} with all changes explained in detail in the sequel.

The pre-processing step consists of two steps including the creation of  ship-centred and masked images and the inverse problem to obtain $\hat{X}_{\text{MAP}}$ as discussed in the previous sections. For the masking operation in Graziano et. al. \cite{graziano1}, instead of replacing the area of the ship location with the mean intensity as is done in this paper, only the unmasked pixels have been taken into account and the area containing the ship and land returns is ignored (for details see p.4 in \cite{graziano1}).

Upon obtaining the estimate $\hat{X}_{\text{MAP}}$, the first detection step is to restrict the peak/trough searching area in the Radon domain ($r, \theta$). Since the image is centered on the ship, we ensure that the peaks/trough of all possible wakes are located between two sine waves \cite{tunaley2003estimation,graziano1} with: $|r| \leq A\sin\theta$ where the sine wave peak point $A$ refers to the maximum azimuth shift. Although antenna velocity and ship velocity along the slant range could be calculated using the slant range in the presence of field data, we chose to select $A$ depending on the number of pixels in the image azimuth dimension $M$.

The value $A$ has a crucial importance in the detection process. Selecting a large value for A increases the searching area, which is increasing the possibility of mis-detecting noisy peaks as wakes. Conversely, selecting it as a small value may cause the ship wakes to fall outside the search area. Here, we use $A=M/10$ which we have shown to lead to the best results for a range of $[M/15, M/5]$ in \cite{karakucs2019ship}.

In the literature, it has been widely stated that the most visible wake is the turbulent wake \cite{eldhuset1996automatic,lyden1988synthetic} and as stated in Section \ref{sec:sec2}, in most cases, it can be surrounded by bright narrow V-wake. Hence, the next step is detecting a peak/trough pair in the restricted area of $\hat{X}_{\text{MAP}}$ that corresponds to the turbulent and one arm of the narrow V-wake, respectively. A window of size $4^{\circ}$ for detecting turbulent/narrow V-wake pair is then scanned. The pair which maximises the difference in amplitudes is selected as turbulent/narrow V-wake pair. Following the detection of the turbulent/narrow V-wake pair, the other arm of the narrow V-wake is searched on the other side of the turbulent wake within the same window size. The point with the maximum amplitude is selected as the second narrow V-wake.

As discussed in the previous sections, Kelvin wake is outer signature of a moving vessel. In order to detect both Kelvin arms, both sides of the detected turbulent wake are searched with a half angle window of size $10^{\circ}$ starting from $\pm10^{\circ}$ to $\pm20^{\circ}$ \cite{reed1990hydrodynamics,zilman2015detectability} and points with maximum amplitudes on each side are then selected as Kelvin arms.

Up to this point, candidate wakes are detected by using the inverse problem solution $\hat{X}_{\text{MAP}}$, and the next step consists in validating the detected wakes, which includes removal of half lines and confirmation of detected wakes via measure indexes. Validation step is performed in image domain since it has been stated in \cite{grazianoradonMerit} that validation in the Radon domain might lead to erroneous results. Hence, points detected in the Radon domain are instead transformed into lines in the image domain via the inverse Radon transform.

The half of the lines are first removed. This has been referred to as the $180^{\circ}$ ambiguity problem in \cite{graziano1}. To solve this ambiguity, only the detected turbulent wake is used. The average intensity over the line representing the turbulent wake is calculated and the half line having lower average (since the turbulent wake is a dark line) is selected as the un-confirmed half line corresponding to the turbulent wake. Half lines belonging to the other detected wakes are selected if they are located in $\pm45^{\circ}$ on either side of the un-confirmed turbulent wake.

Confirmation of the candidate half lines is then performed using a measure index $F_I$, which is calculated as: $F_I = \bar{I}_w/\bar{I} - 1$ and interpreted as a measure of the difference between the average intensity over the un-confirmed wake, $\bar{I}_w$, and the average intensity of the image, $\bar{I}$. The $F_I$ index is positive for bright wakes and negative only for the turbulent wake. Moreover, deciding a margin will help to reduce the possibility of false confirmations. Thus, we assumed a margin of 10\% after a trial-error procedure (Please see Section \ref{sec:simImages}). Detected half lines which do not follow:
\begin{align}\label{equ:index}\begin{tabular}{ll}
                               $F_I < 0$  & for turbulent wake, \\
                               $F_I > 0.1$  & for narrow-V and Kelvin wakes. \\
                              \end{tabular}
\end{align}
are discarded, whereas the remaining half lines are confirmed.

\section{SAR Data Sets}\label{sec:Data}
In this section, we describe the datasets employed in the experimental part of the paper. These consist of both simulated and real SAR images.

\subsection{Simulated SAR Images of The Sea Surface}\label{sec:SimSar}
In order to generate simulated SAR images of the sea surface, a numerical SAR image simulation software was developed based on previous studies in \cite{zilman2015detectability, oumansour1996multifrequency, lyzenga1986numerical}. The first part of the simulations consists in sea surface modelling, where the linear theory of surface waves was used. The ship wake modelling (Kelvin wake) considers ships as rigid bodies moving in inviscid incompressible fluid. Here, basic parameters belonging to ship such as length, beam, draft and the Froude number are used to model different types of wakes.

The SAR imaging mechanism is usually described via real aperture radar (RAR) and specific SAR imaging. The RAR imaging is represented as a normalised radar cross section (NRCS) backscattering with vertical-vertical (VV) and horizontal-horizontal (HH) signal polarisation modes and two linear modulations of tilt and hydrodynamic. Specific SAR imaging is based on a velocity bunching modulation (For details see \cite{zilman2015detectability,zurk1996comparison} and references therein).

\subsection{Real SAR Data}\label{sec:RealSar}
SAR images from four different satellite platforms, namely TerraSAR-X, COSMO-SkyMed, Sentinel-1 and ALOS2 are employed. All TerraSAR-X images \cite{terraSARX} are X-band Stripmap products with three meters resolution for both azimuth and range directions. We have used 3 images, two of which are in HH polarisation with the third in VV polarisation. Of the COSMO-SkyMed data sets, we have used 2 X-band images including ship wake patterns. These are Stripmap products and their resolution is three meters for both azimuth and range directions, in different polarisations. The Sentinel-1 images utilised in this paper are C-band SAR images and have ten meters resolution for both azimuth and range directions. We have used 4 images, all of which are in VV polarisation. Lastly, we have used 2 ALOS2 images which are L-band and in HH polarisation. ALOS2 images have three meters resolution. From all images, we have selected 28 different ship wake patterns (28 different ships with corresponding wakes) by visual inspection and used them for experimental analysis.

We created ship centered image tiles and used them as input for the ship wake detection procedure discussed in the previous section. We assume that ship locations for the selected ships are known. In practice, this step can be replaced with ship detection techniques such as constant false alarm rate (CFAR) approaches (see e.g. \cite{pappas2018superpixel}).

\begin{table*}[htbp]
  \centering
  \caption{Details for Real SAR Images.}\renewcommand{\arraystretch}{0.9}
    \resizebox{\linewidth}{!}{\begin{tabular}{C{0.11\textwidth}C{0.11\textwidth}C{0.11\textwidth}C{0.11\textwidth}C{0.11\textwidth}C{0.055\textwidth}C{0.055\textwidth}C{0.055\textwidth}C{0.055\textwidth}C{0.055\textwidth}C{0.055\textwidth}C{0.055\textwidth}C{0.055\textwidth}}
    \hline
          &       &   &    &       &       &  \multicolumn{1}{l}{\textbf{Wake}}     &       & \multicolumn{5}{c}{\textbf{Visible Wakes}} \\
    \multicolumn{1}{l}{\textbf{Image}} & \multicolumn{1}{l}{\textbf{Satellite}} & \multicolumn{1}{l}{\textbf{Frequency Band}}&\multicolumn{1}{c}{\textbf{Polarization}} & \multicolumn{1}{l}{\textbf{Resolution}} &       & \multicolumn{1}{l}{\textbf{Number}} &       & \multicolumn{1}{l}{\textbf{Turbulent}} & \multicolumn{1}{l}{\textbf{Narrow1}} & \multicolumn{1}{l}{\textbf{Narrow2}} & \multicolumn{1}{l}{\textbf{Kelvin1}} & \multicolumn{1}{l}{\textbf{Kelvin2}} \\
    \hline
    \multicolumn{1}{l}{Image 1} & \multicolumn{1}{l}{TerraSAR-X} & \multicolumn{1}{l}{X (8-12 GHz)} &\multicolumn{1}{c}{HH} & \multicolumn{1}{l}{$3\times3$ m} &       & 1.1     &       & \cmark     & \cmark     & \cmark     & \cmark     & \cmark \\
          &       &    &   &       &       & 1.2     &       & \cmark     & \cmark     & \xmark     & \cmark     & \xmark \\
          &       &    &   &       &       & 1.3     &       & \cmark     & \cmark     & \xmark     & \cmark     & \xmark \\
          &       &    &   &       &       & 1.4     &       & \cmark     & \cmark     & \xmark     & \xmark     & \xmark \\
          \hline
    \multicolumn{1}{l}{Image 2} & \multicolumn{1}{l}{TerraSAR-X} & \multicolumn{1}{l}{X (8-12 GHz)} &\multicolumn{1}{c}{HH} & \multicolumn{1}{l}{$3\times3$ m} &       & 2.1     &       & \cmark     & \cmark    & \xmark     & \cmark     &\cmark\\
          &       &    &   &       &       & 2.2     &       & \cmark     & \cmark     & \xmark     & \xmark     & \xmark \\
          &       &    &   &       &       & 2.3     &       &\cmark     & \cmark     & \xmark     & \cmark     & \xmark \\
          &       &    &   &       &       & 2.4     &       &\cmark     & \cmark     & \xmark     & \cmark     & \xmark \\
          &       &    &   &       &       & 2.5     &       &\cmark     & \cmark     & \cmark     & \cmark     & \cmark \\
          \hline
    \multicolumn{1}{l}{Image 3} & \multicolumn{1}{l}{TerraSAR-X} & \multicolumn{1}{l}{X (8-12 GHz)} &\multicolumn{1}{c}{VV} & \multicolumn{1}{l}{$3\times3$ m} &       & 3.1     &       & \cmark     & \cmark     & \xmark     & \cmark     & \xmark \\
    &       &   &    &       &       & 3.2     &       & \cmark     & \cmark     & \xmark     & \cmark     & \xmark \\
     &       &   &    &       &       & 3.3     &       & \cmark     & \cmark     & \xmark     & \cmark     & \xmark \\
          \hline
    \multicolumn{1}{l}{Image 4} & \multicolumn{1}{l}{ALOS2} & \multicolumn{1}{l}{L (1-2 GHz)} &\multicolumn{1}{c}{HH} & \multicolumn{1}{l}{$3\times3$ m} &       & 4.1     &       & \cmark     & \cmark     & \xmark     & \xmark     & \xmark \\
          \hline
    \multicolumn{1}{l}{Image 5} & \multicolumn{1}{l}{ALOS2} & \multicolumn{1}{l}{L (1-2 GHz)} &\multicolumn{1}{c}{HH} & \multicolumn{1}{l}{$3\times3$ m} &       & 5.1    &       & \cmark     & \cmark     & \xmark     & \xmark     & \xmark \\
    &       &   &    &       &       & 5.2     &       & \cmark     & \cmark     & \xmark     & \xmark     & \xmark \\
    &       &   &    &       &       & 5.3     &       & \cmark     & \cmark     & \xmark     & \cmark     & \xmark \\
          \hline
          \multicolumn{1}{l}{Image 6} & \multicolumn{1}{l}{COSMO-SkyMed} & \multicolumn{1}{l}{X (8-12 GHz)} &\multicolumn{1}{c}{VV} & \multicolumn{1}{l}{$3\times3$ m} &       & 6.1     &       & \cmark     & \cmark    & \cmark     & \cmark     &\xmark\\
          &       &   &    &       &       & 6.2     &       & \cmark     & \cmark    & \xmark     & \xmark     &\xmark\\
     &       &   &    &       &       & 6.3     &       & \cmark     & \cmark    & \xmark     & \cmark     &\xmark\\
     \hline
                    \multicolumn{1}{l}{Image 7} & \multicolumn{1}{l}{COSMO-SkyMed} & \multicolumn{1}{l}{X (8-12 GHz)} &\multicolumn{1}{c}{HH} & \multicolumn{1}{l}{$3\times3$ m} &       & 7.1    &       & \cmark     & \cmark    & \xmark     & \cmark     &\xmark\\
          \hline
          \multicolumn{1}{l}{Image 8} & \multicolumn{1}{l}{Sentinel-1} & \multicolumn{1}{l}{C (4-8 GHz)} &\multicolumn{1}{c}{VV} & \multicolumn{1}{l}{$10\times10$ m} &       & 8.1     &       & \cmark     & \cmark    & \xmark     & \xmark     &\xmark\\
          &       &    &   &       &       & 8.2     &       & \cmark     & \cmark     & \xmark     & \cmark     & \xmark \\
          &       &    &   &       &       & 8.3     &       & \cmark     & \cmark     & \xmark     & \xmark     & \xmark \\
          \hline
                    \multicolumn{1}{l}{Image 9} & \multicolumn{1}{l}{Sentinel-1} & \multicolumn{1}{l}{C (4-8 GHz)} &\multicolumn{1}{c}{VV} & \multicolumn{1}{l}{$10\times10$ m} &       & 9.1    &       & \cmark     & \cmark    & \xmark     & \xmark     &\xmark\\
          \hline
          \multicolumn{1}{l}{Image 10} & \multicolumn{1}{l}{Sentinel-1} & \multicolumn{1}{l}{C (4-8 GHz)} &\multicolumn{1}{c}{VV} & \multicolumn{1}{l}{$10\times10$ m} &       & 10.1     &       & \cmark     & \cmark     & \xmark     & \xmark     & \xmark \\
          &       &   &    &       &       & 10.2     &       &\cmark     & \cmark     & \xmark     & \xmark     & \xmark \\
          &       &   &    &       &       & 10.3     &       & \cmark     & \cmark     & \xmark     & \xmark     & \xmark \\
          \hline
                    \multicolumn{1}{l}{Image 11} & \multicolumn{1}{l}{Sentinel-1} & \multicolumn{1}{l}{C (4-8 GHz)} &\multicolumn{1}{c}{VV} & \multicolumn{1}{l}{$10\times10$ m} &       & 11.1     &       & \cmark     & \cmark    & \xmark     & \xmark     &\xmark\\
          \hline
    \end{tabular}}%
  \label{tab:visibleWakes}%
\end{table*}%

All ships in the created ship-centered image tiles are first masked to remove the bright spots in the Radon domain resulting from the ship itself. This allows us to discriminate bright points as possible ship wakes in the Radon domain. Following the pre-processing operations, the proposed inverse problem based ship wake detection algorithm has been tested using the 28 aforementioned image tiles. In Table \ref{tab:visibleWakes}, each image tile is shown with their visible wakes and corresponding details. As it can be seen, only 2 image tiles (1.1 and 2.6) out of 28 have all five possible types of ship wakes. The remaining images have less than five wakes and the performance of the proposed method and the reference methods are evaluated as true detections of ship wakes. True detection in experimental analysis implies detecting visible wakes, and discarding invisible wakes.

\section{Experimental Results}\label{sec:Exp}
The proposed method was tested from three different perspectives using both simulated and real data. i) We first used simulated SAR images of the sea surface including ship wakes.  ii) Subsequently, we conducted experiments to determine the best choice of prior for solving the inverse problem in eq. (\ref{equ:invprob}). iii) Lastly, we compared the best method tested at (ii) to state of the art approaches for ship wake detection. The performance comparison was carried out in terms of receiver operation characteristics (ROC) as well as other measures, which are specifically described in Table \ref{tab:perfMetrics}.

Sensitivity quantifies the number of true positive in proportion to the total number of detections and specificity does the same for true negatives. The percentage accuracy shows the correct detection percentage over TP and TN values and was used to assess the ship wake detection performance of the method directly. The $F_1$ score is also a metric which shows the accuracy of the methods. Positive likelihood ratio (LR+) shows how likely (or suitable) the utilised model is to detect correct classes. The last metric used in this paper is Youden's $J$ index, which shows the success of the test with values between 0 and 1. The value 0 implies the ``test is unsuccessful'' whereas 1 implies ``the test is successful''.

\begin{table}[htbp]
  \centering
  \caption{Descriptions OF all parameters, priors and performance comparison metrics.}
    \resizebox{.9\linewidth}{!}{\begin{tabular}{lcl}
    \hline
    Expression && Description \\
    \hline
    $M$ && Image dimension. ($M\times M$)\\
    $k$ && Parameter defines GMC, $L_1$, $L_p$, TV and nuclear priors for \\
    && 1, 2, 3, 4 and 5, respectively.\\
    $\lambda_k$ && Regularisation constant for prior $k$.\\
    $\gamma$ && Parameter for GMC which controls convexity. Set to 0.9.\\
    $\epsilon^{(i)}$ && Error criterion term for iteration $i$ in Algorithm \ref{alg:FB2}. Refer to\\
    && (\ref{equ:Stop})\\
    $MaxIter$ && Maximum number of iterations for Algorithms \ref{alg:FB} and \ref{alg:FB2}.\\
    && Set to 1000.\\
    $\psi_k(X)$ && Negative logarithm of prior $k$. Please refer to (\ref{equ:GMCprior}), (\ref{equ:L1prior2}), (\ref{equ:Lpprior2}),\\
    && (\ref{equ:TVprior3}) and (\ref{equ:Nucprior2}) for $k=1, 2, 3, 4, 5$, respectively.\\
    $p$ && $L_p$ norm order. $0 < p < 1$.\\
    $A$ && Maximum sine wave amplitude for wake detection process.\\
    && Set to $M/10$. Refer to Section \ref{sec:Detection}.\\
    Kelvin wake searching range && An interval of $\pm[10^o, 20^o]$ on either side of turbulent wake.\\
    Narrow V-wake searching range && An interval of $\pm[0^o, 4^o]$ on either side of turbulent wake.\\
    True Positive (TP) && Correct confirmation of visible wakes. \\
    True Negative (TN) && Correct discard of invisible wakes \\
    False Positive (FP) && False detection of invisible wakes \\
          && FP includes mislocated wake confirmations. \\
    False Negative (FN) && False discard of visible ship wakes.  \\
    N && Total number of all possible wakes.\\
    &&Equal to 140 (28 Images $\times$ 5 possible wakes)\\
    Sensitivity && TP/(TP+FN) \\
    Specificity && TN/(TN + FP) \\
    \% Accuracy && 100(TP+TN)/N \\
    $F_1$ && 2TP/(2TP + FP + FN)\\
    LR+ && Sensitivity/(1-Specificity)\\
    Youden's $J$ && Sensitivity + Specificity - 1\\
    \hline
    \end{tabular}}%
  \label{tab:perfMetrics}%
\end{table}%

\subsection{Wake detection in simulated SAR images of sea surface}\label{sec:simImages}
In the first set of experiments, we tested the proposed method with simulated SAR images of the sea surface. We generated two SAR images with VV polarisation and a 4 m/s wind speed, 50 m size of ship, which is moving with velocity of 9 m/s. In addition, both images are with different ship orientation, which effects the visibility of the wakes \cite{panico2017sar}. Inverse problem solution with all employed priors are obtained and the procedure described above for detecting ship wakes is modified just to detect Kelvin arms instead of all five wakes as simulated images only contain 1 or 2 visible Kelvin arms.

\begin{figure*}[ht!]
\centering
\subfigure[Simulated Image]{
\includegraphics[width=.23\linewidth]{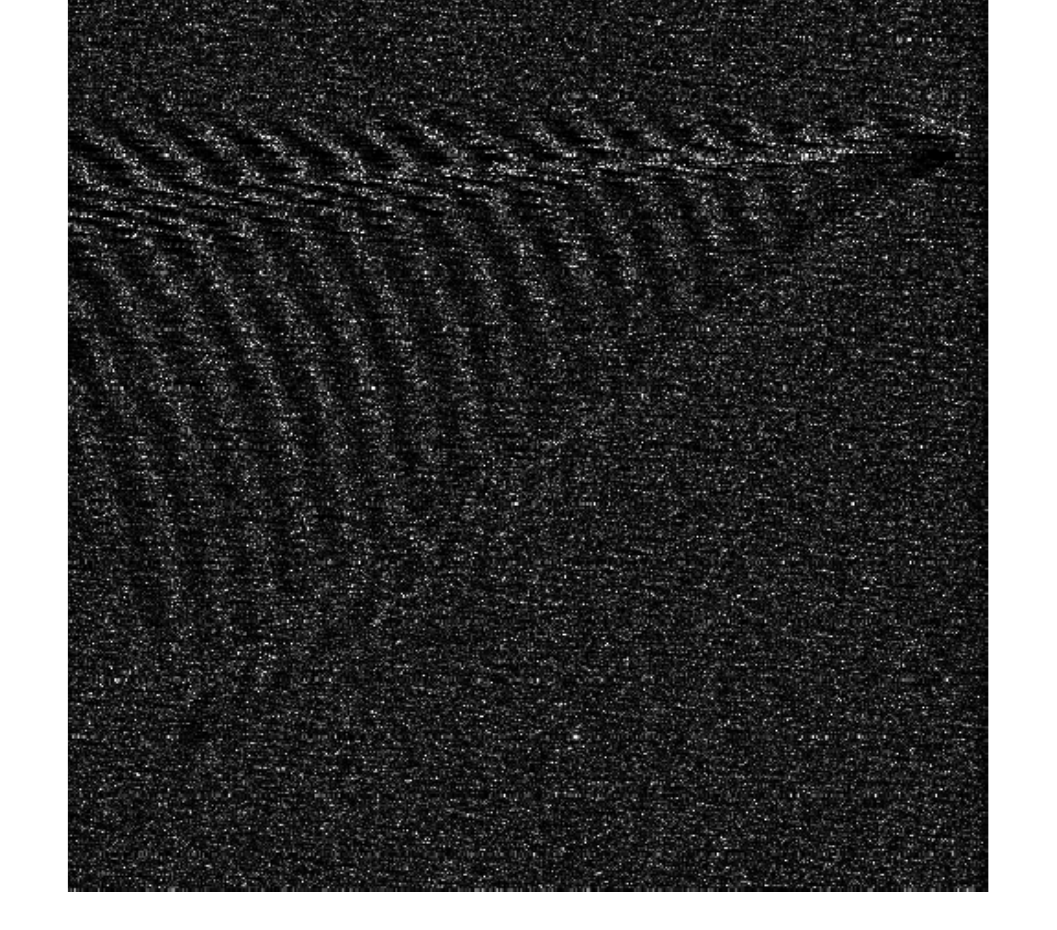}}
\centering
\subfigure[GMC]{
\includegraphics[width=.23\linewidth]{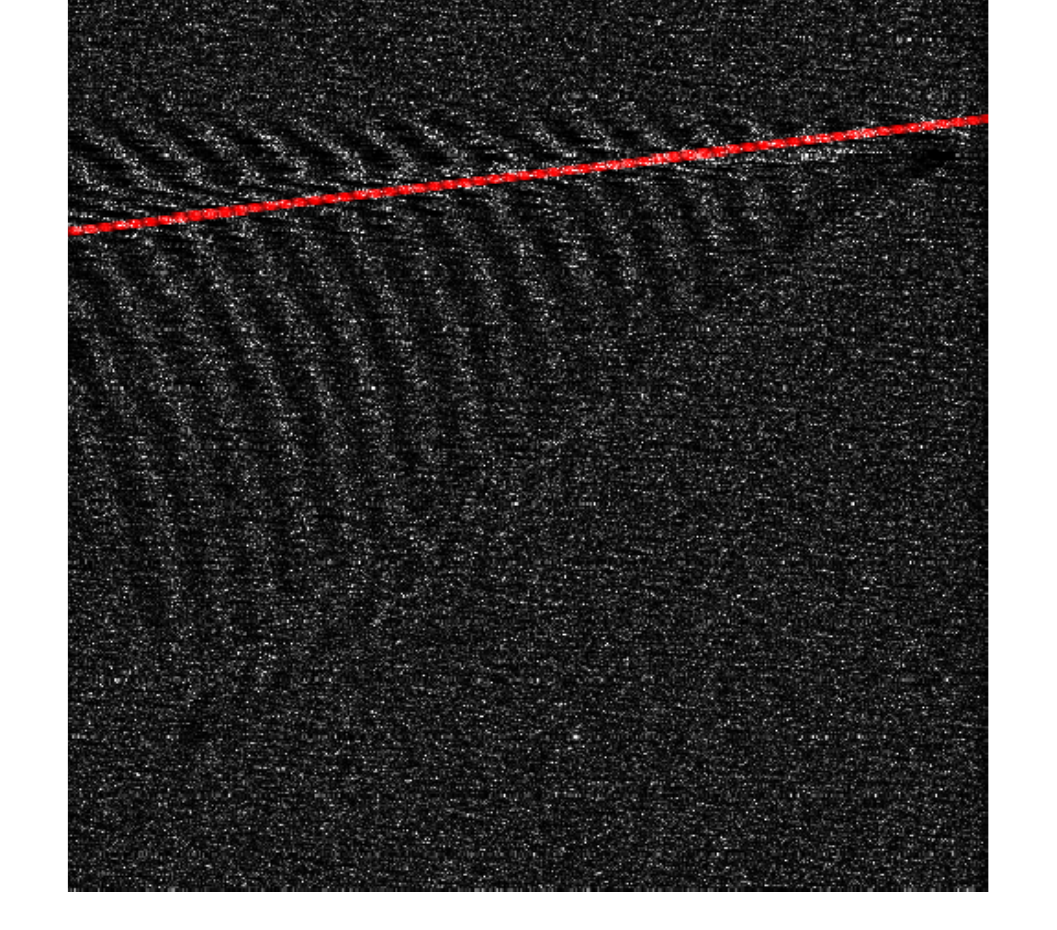}}
\centering
\subfigure[$L_1$]{
\includegraphics[width=.23\linewidth]{simSWD.pdf}}
\centering
\subfigure[TV]{
\includegraphics[width=.23\linewidth]{simSWD.pdf}}
\centering
\subfigure[Simulated Image]{
\includegraphics[width=.23\linewidth]{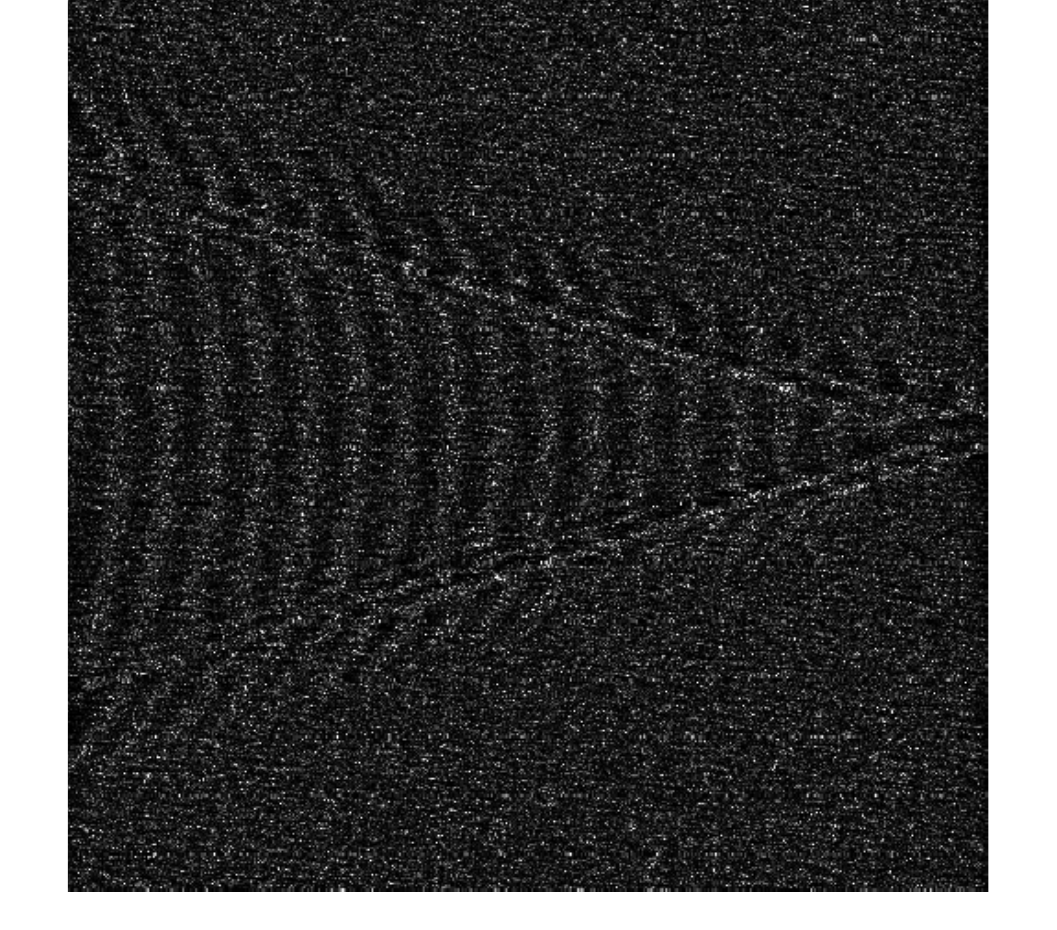}}
\centering
\subfigure[GMC]{
\includegraphics[width=.23\linewidth]{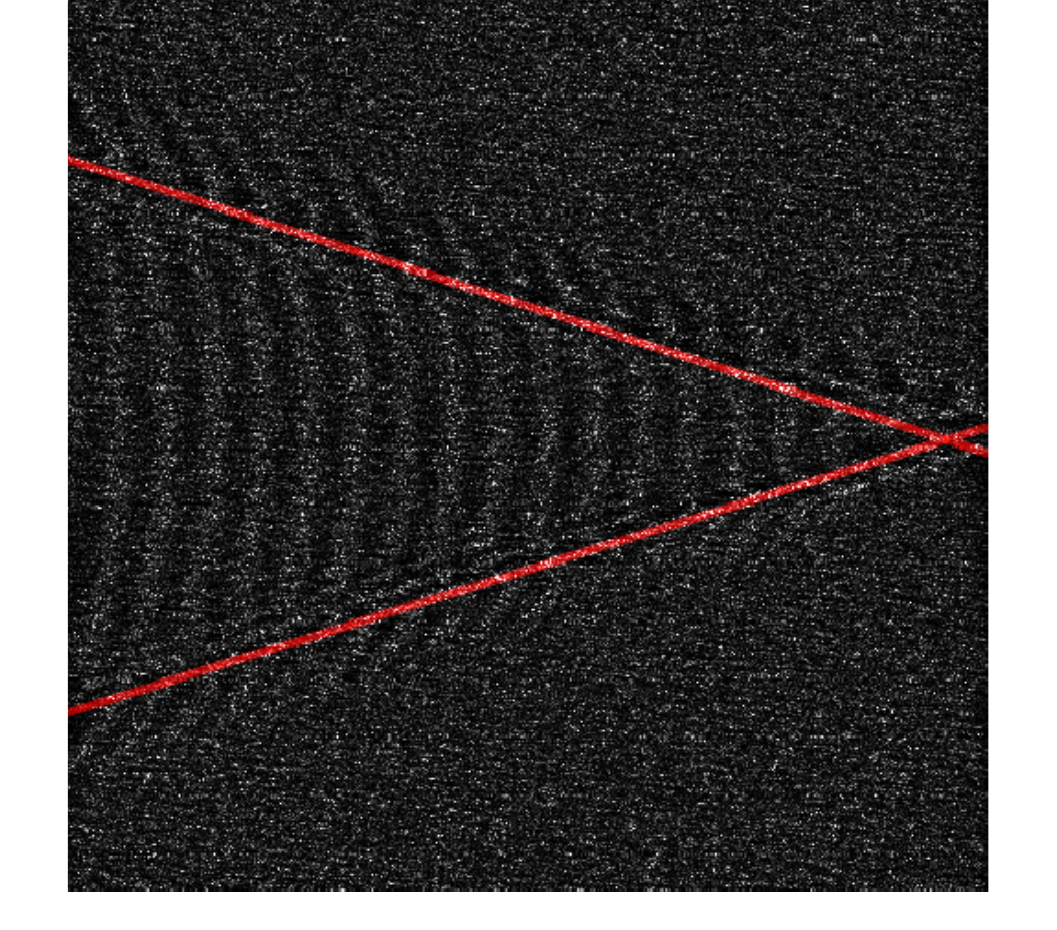}}
\centering
\subfigure[$L_1$]{
\includegraphics[width=.23\linewidth]{simSWD2.pdf}}
\centering
\subfigure[TV]{
\includegraphics[width=.23\linewidth]{simSWD2.pdf}}
\caption{Ship wake detection for simulated SAR images of the sea surface. (a), (e) Simulated images, (b)-(d) and (f)-(h) Ship wake detection results after $F>0.1$ confirmation step.}\vspace{-0.3cm}
\label{fig:simImage}
\end{figure*}

A comparison study has been carried out for both images and all visible wakes are successfully detected for all types of priors. Moreover, we apply different $F$ index values to determine the most suitable value for $F$ in the confirmation step and we conclude that the $F$ index value of 0.1 is enough to discard wrong detections which follows our assumption for real SAR images in this paper. In Figure \ref{fig:simImage}, simulated images and the wake detection results for GMC, $L_1$ and TV are depicted. On examining the figure, it is clear that all three visible Kelvin wakes are detected and the invisible one in Figure \ref{fig:simImage}-(a) is discarded.

\subsection{Inverse problem solution for various priors}
In the second set of experiments, we tested the proposed ship wake detection framework in the context of using different types of priors. The results are presented in Table \ref{tab:normComp}. The performance metrics in Table \ref{tab:normComp} corresponds to all 28 real SAR images in our data set. Examining the values in Table \ref{tab:normComp}, it is obvious to state that the inverse problem approach based on the GMC achieves the highest performance metrics among all methods tested. The percentage accuracy is around 10\% higher than the TV, which is the second best.
\begin{figure}[ht!]
\centering
\includegraphics[width=.65\linewidth]{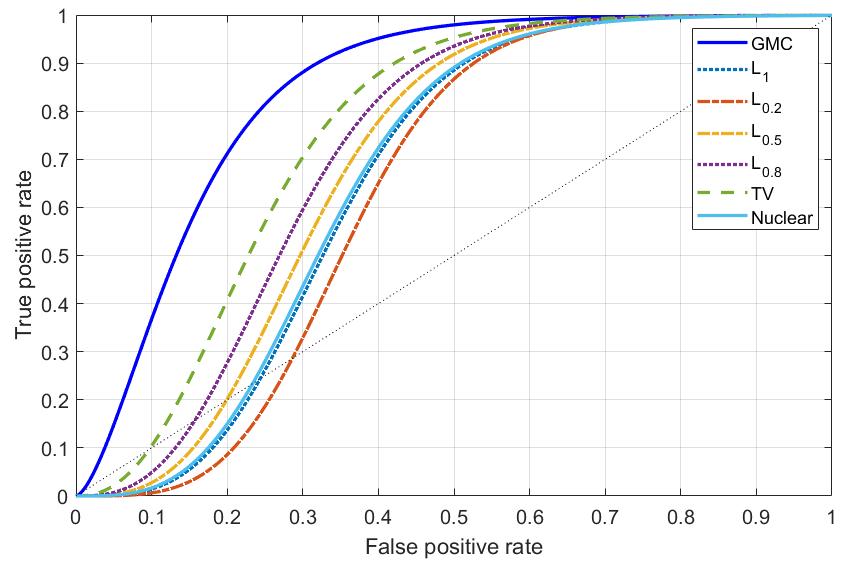}
\caption{Performance comparison of different priors in terms of a receiver operating characteristics (ROC) curve.}
\label{fig:rocNorms}
\end{figure}

\begin{table*}[htbp]
  \centering
  \caption{Detection performance of different priors over all data sets.}
    \resizebox{.84\linewidth}{!}{
    \begin{tabular}{cccccccccccccccccccccccccccccccccccccccc}
    \hline
          &       &       &       & \multicolumn{4}{c}{TP}        & \multicolumn{4}{c}{TN}        & \multicolumn{4}{c}{FP}        & \multicolumn{4}{c}{FN}        & \multicolumn{4}{c}{Sensitivity} & \multicolumn{4}{c}{Specificity} & \multicolumn{3}{|c}{\% Accuracy} & \multicolumn{3}{c}{$F_1$} & \multicolumn{3}{c}{LR+} & \multicolumn{3}{c}{Youden's $J$} \\
          \hline
    \multicolumn{4}{c}{GMC}       & \multicolumn{4}{c}{49.29\%}   & \multicolumn{4}{c}{30.71\%}   & \multicolumn{4}{c}{17.86\%}   & \multicolumn{4}{c}{2.14\%}    & \multicolumn{4}{c}{95.83\%}   & \multicolumn{4}{c|}{63.24\%}  & \multicolumn{3}{c}{\textbf{80.00\%}} & \multicolumn{3}{c}{\textbf{0.83}} & \multicolumn{3}{c}{\textbf{2.61}} & \multicolumn{3}{c}{\textbf{0.59}} \\
    \multicolumn{4}{c}{$L_1$}        & \multicolumn{4}{c}{35.00\%}   & \multicolumn{4}{c}{28.57\%}   & \multicolumn{4}{c}{32.14\%}   & \multicolumn{4}{c}{4.29\%}    & \multicolumn{4}{c}{89.09\%}   & \multicolumn{4}{c|}{47.06\%}  & \multicolumn{3}{c}{63.57\%} & \multicolumn{3}{c}{0.66} & \multicolumn{3}{c}{1.68} & \multicolumn{3}{c}{0.36} \\
    \multicolumn{4}{c}{$L_{02}$}       & \multicolumn{4}{c}{40.00\%}   & \multicolumn{4}{c}{26.43\%}   & \multicolumn{4}{c}{32.86\%}   & \multicolumn{4}{c}{0.71\%}    & \multicolumn{4}{c}{98.25\%}   & \multicolumn{4}{c|}{44.58\%}  & \multicolumn{3}{c}{66.43\%} & \multicolumn{3}{c}{0.70} & \multicolumn{3}{c}{1.77} & \multicolumn{3}{c}{0.43} \\
    \multicolumn{4}{c}{$L_{05}$}       & \multicolumn{4}{c}{37.86\%}   & \multicolumn{4}{c}{30.00\%}   & \multicolumn{4}{c}{30.71\%}   & \multicolumn{4}{c}{1.43\%}    & \multicolumn{4}{c}{96.36\%}   & \multicolumn{4}{c|}{49.41\%}  & \multicolumn{3}{c}{67.86\%} & \multicolumn{3}{c}{0.70} & \multicolumn{3}{c}{1.90} & \multicolumn{3}{c}{0.46} \\
    \multicolumn{4}{c}{$L_{08}$}       & \multicolumn{4}{c}{33.57\%}   & \multicolumn{4}{c}{32.14\%}   & \multicolumn{4}{c}{29.29\%}   & \multicolumn{4}{c}{5.00\%}    & \multicolumn{4}{c}{87.04\%}   & \multicolumn{4}{c|}{52.33\%}  & \multicolumn{3}{c}{65.71\%} & \multicolumn{3}{c}{0.66} & \multicolumn{3}{c}{1.83} & \multicolumn{3}{c}{0.39} \\
    \multicolumn{4}{c}{TV}        & \multicolumn{4}{c}{35.71\%}   & \multicolumn{4}{c}{34.29\%}   & \multicolumn{4}{c}{26.43\%}   & \multicolumn{4}{c}{3.57\%}    & \multicolumn{4}{c}{90.91\%}   & \multicolumn{4}{c|}{56.47\%}  & \multicolumn{3}{c}{70.00\%} & \multicolumn{3}{c}{0.70} & \multicolumn{3}{c}{2.09} & \multicolumn{3}{c}{0.47} \\
    \multicolumn{4}{c}{Nuclear}   & \multicolumn{4}{c}{33.57\%}   & \multicolumn{4}{c}{29.29\%}   & \multicolumn{4}{c}{32.14\%}   & \multicolumn{4}{c}{5.00\%}    & \multicolumn{4}{c}{87.04\%}   & \multicolumn{4}{c|}{47.67\%}  & \multicolumn{3}{c}{62.86\%} & \multicolumn{3}{c}{0.64} & \multicolumn{3}{c}{1.66} & \multicolumn{3}{c}{0.35} \\
    \hline
    \end{tabular}}%
  \label{tab:normComp}%
\end{table*}%

ROC curves for all priors are depicted in Figure \ref{fig:rocNorms} which demonstrates ship wake detection performance via true positive rate (TPR, or Sensitivity) vs. false positive rate (FPR, or 1 - Specificity). Examining the plots in Figure \ref{fig:rocNorms}, it can apparently be seen that the proposed GMC based inverse problem method outperforms all other priors in terms of ROC analysis.

\begin{figure*}[ht!]
\centering
\subfigure[Ship centered tile]{
\includegraphics[width=.24\linewidth]{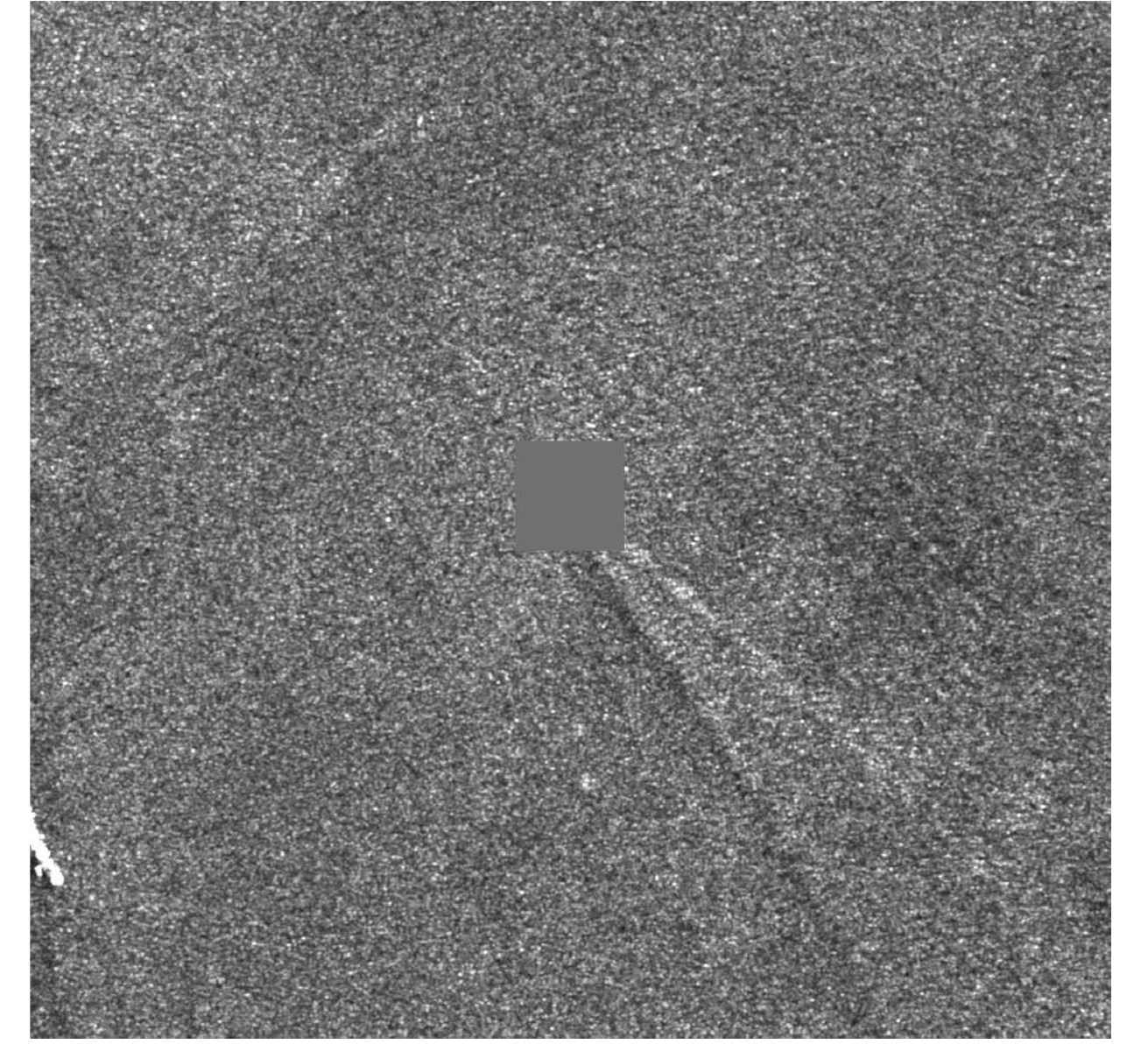}}
\centering
\subfigure[GMC]{
\includegraphics[width=.24\linewidth]{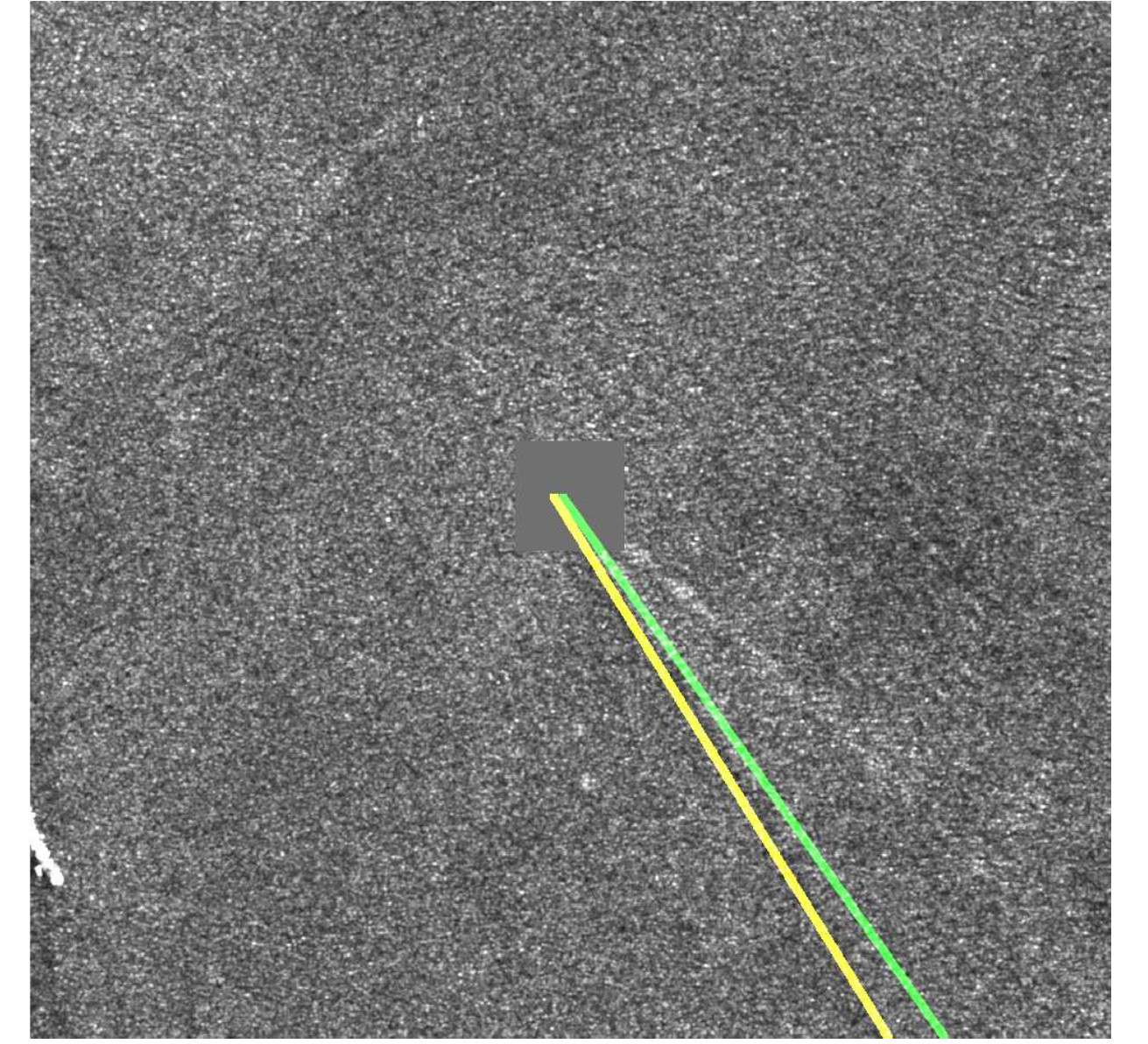}}
\centering
\subfigure[$L_{1}$]{%
\includegraphics[width=.24\linewidth]{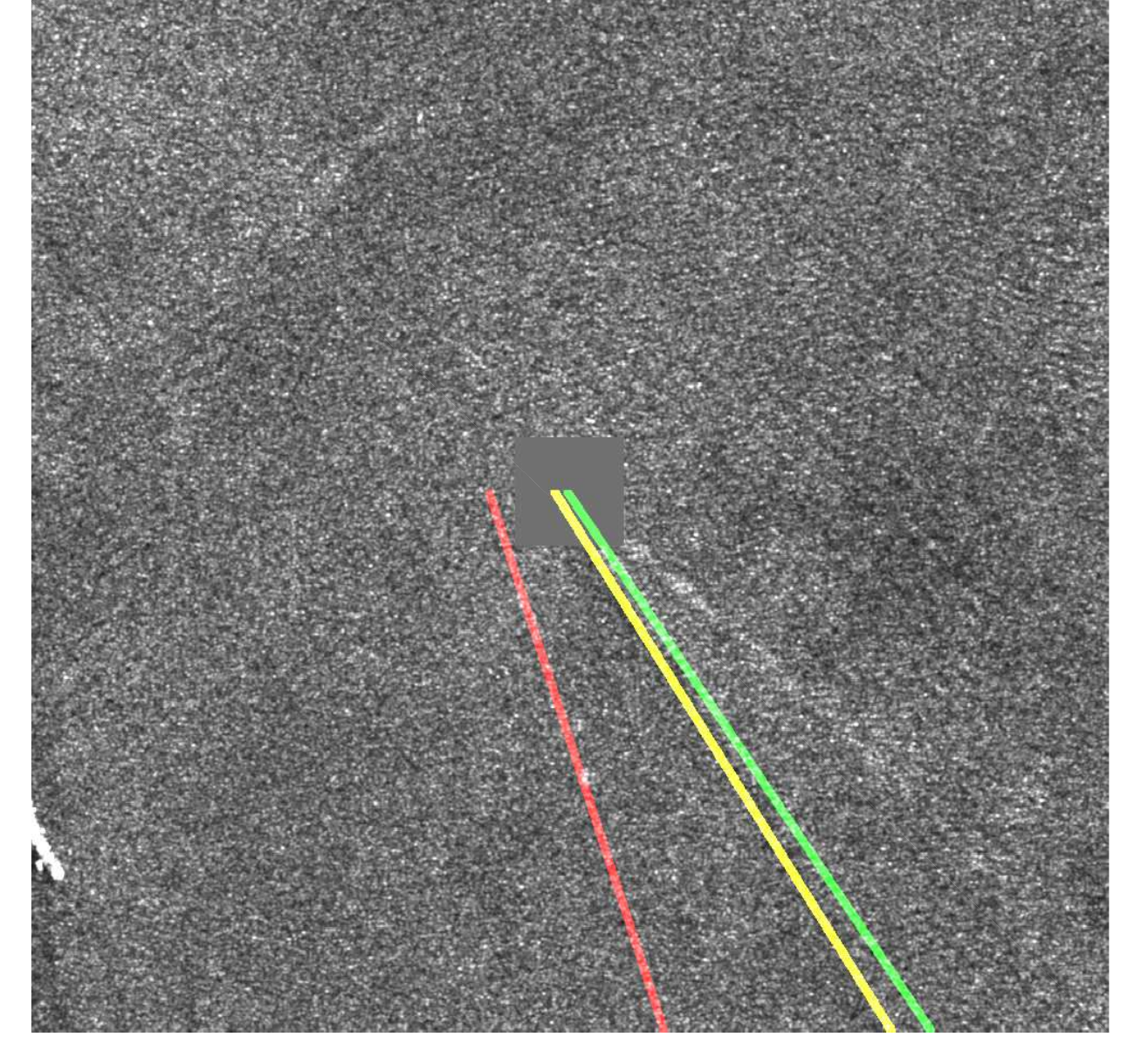}}
\centering
\subfigure[$L_{0.2}$]{%
\includegraphics[width=.24\linewidth]{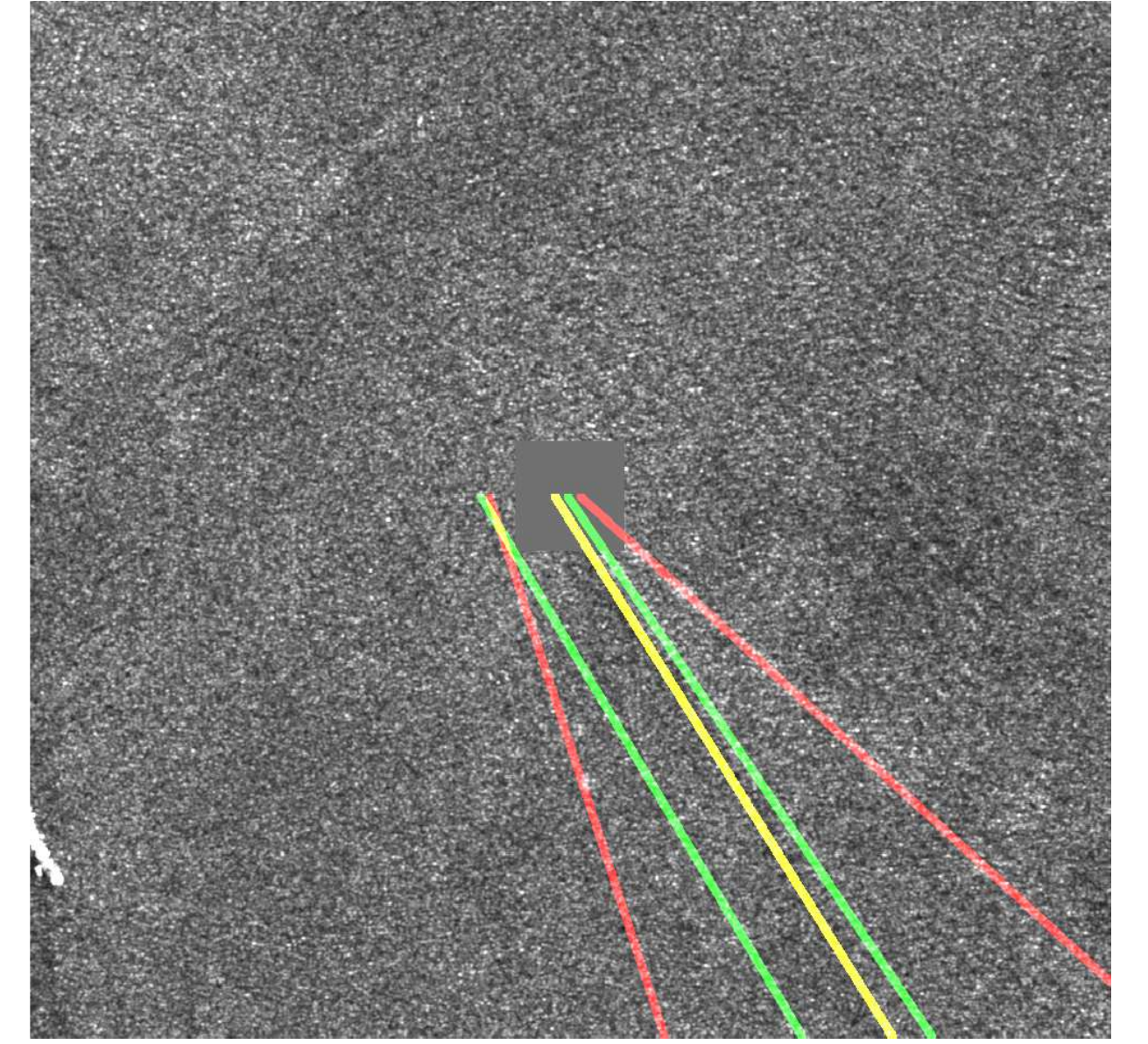}}
\centering
\subfigure[$L_{0.5}$]{%
\includegraphics[width=.24\linewidth]{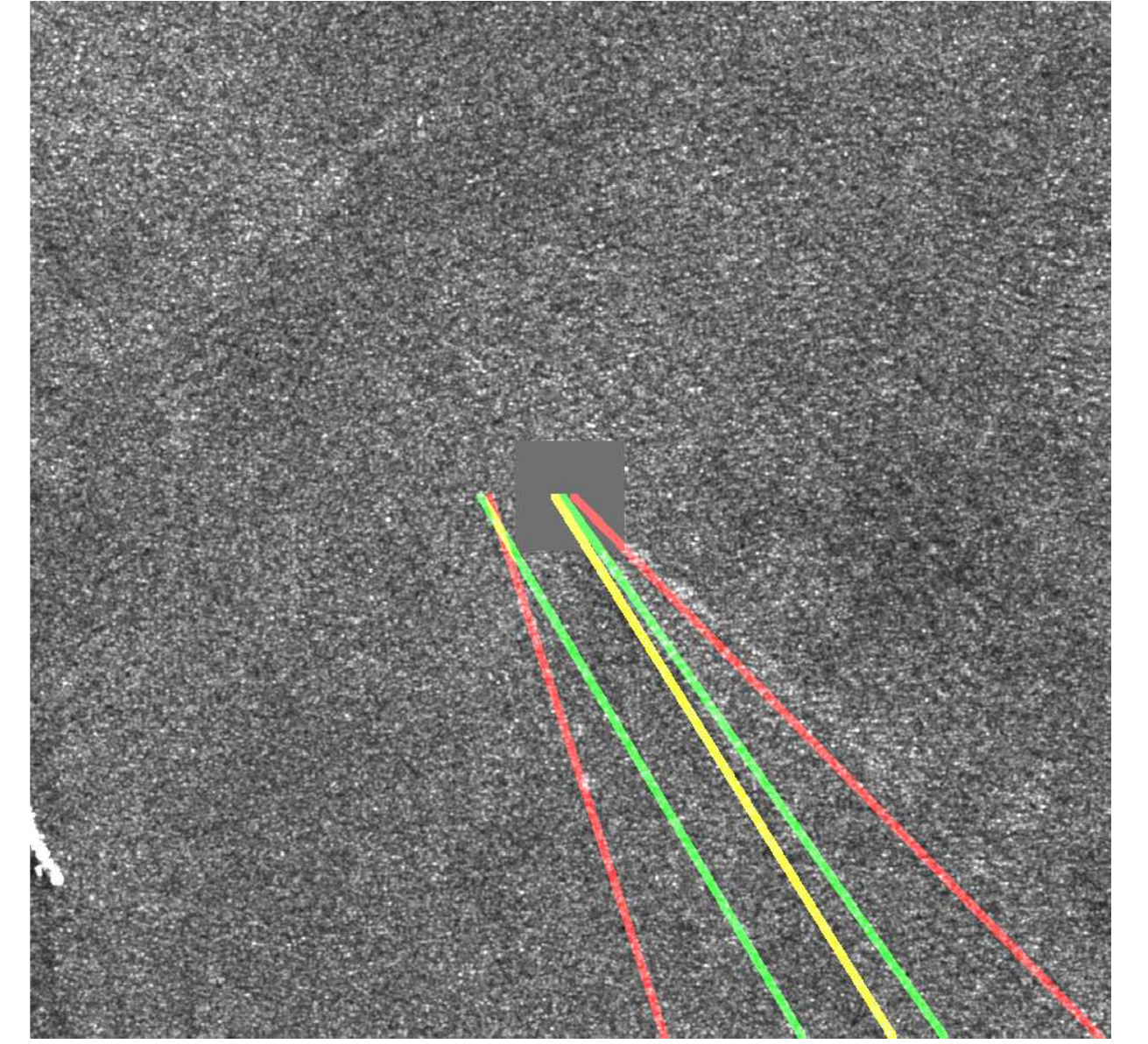}}
\centering
\subfigure[$L_{0.8}$]{%
\includegraphics[width=.24\linewidth]{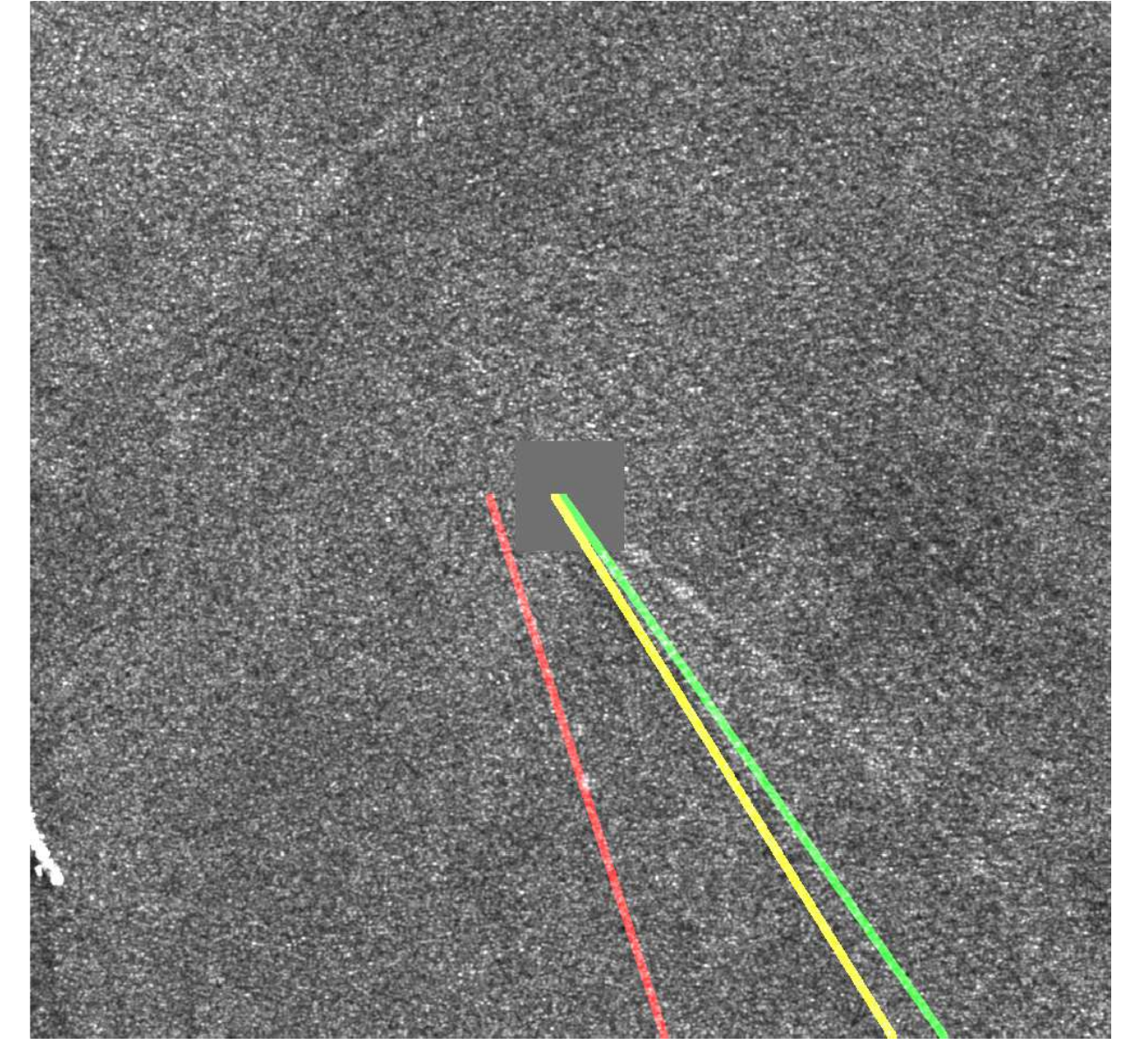}}
\centering
\subfigure[TV]{%
\includegraphics[width=.24\linewidth]{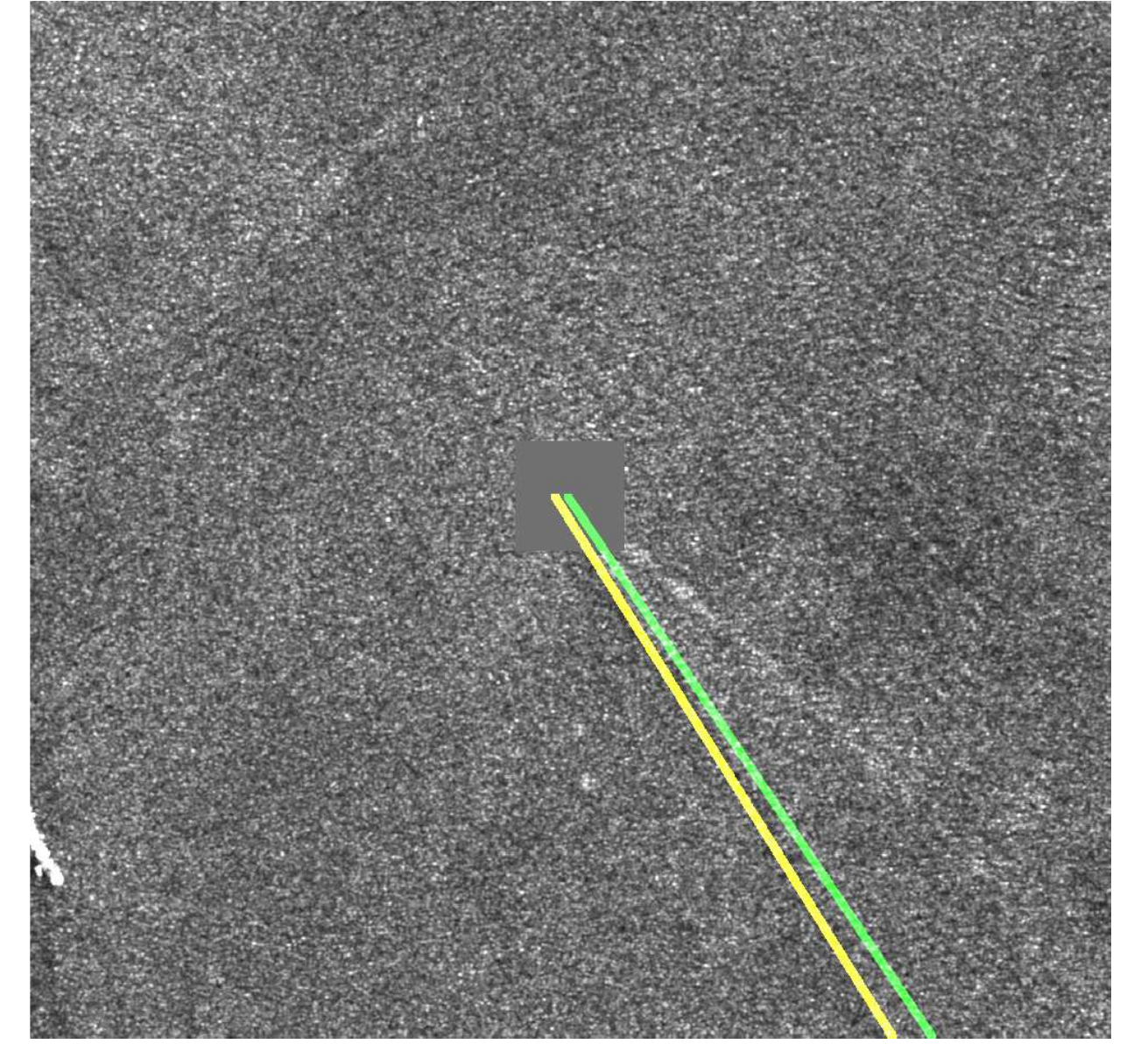}}
\centering
\subfigure[Nuclear]{%
\includegraphics[width=.24\linewidth]{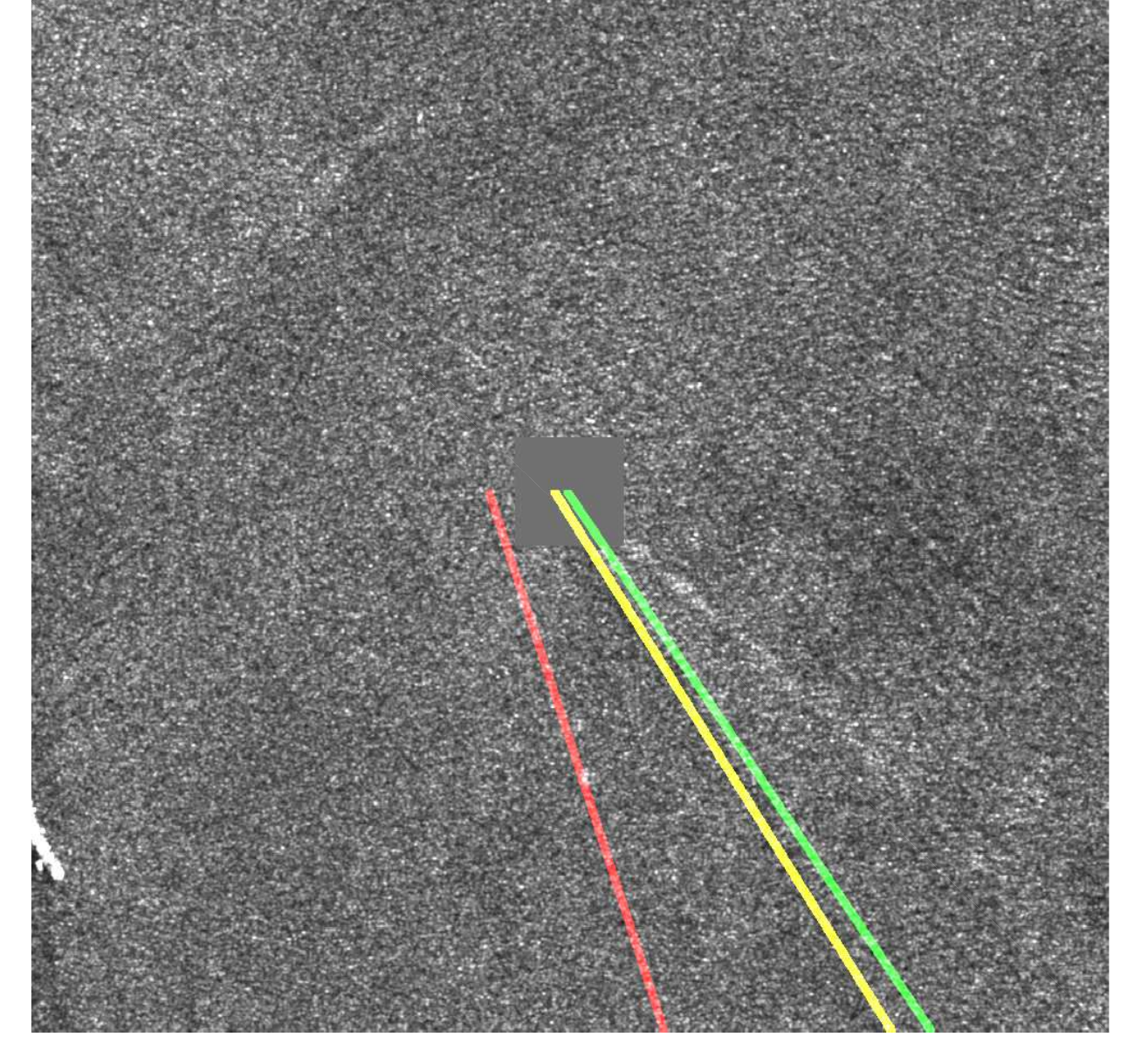}}
\caption{Visual results for ship detection for different priors for Wake 2.2. Yellow, green and red lines represent turbulent, narrow-V and Kelvin wakes, respectively. Ship wake image in (a) has only 3 visible wakes, which are turbulent, one narrow-V and one Kelvin wakes. Detecting these three wakes and discarding two un-imaged wakes at the same time has 100\% accuracy.}\vspace{-0.3cm}
\label{fig:normDetection}
\end{figure*}

In Figure \ref{fig:normDetection}, we show visual results to assess the ship wake detection performance. In Figure \ref{fig:normDetection}-(a) the image for wake 2.2 is shown. There are three visually detected wakes corresponding to turbulent wake, one arm of narrow V-wake and one Kelvin arm. Specifically, in Figure \ref{fig:normDetection}, for wake image 2.2, GMC and TV detect two out of three visible wakes and discard all invisible wakes, which leads to an 80\% detection accuracy. For the rest of the priors, detection accuracy is lower, e.g. for $L_{0.5}$ and $L_1$ it is 60\%.

\subsection{Comparison to the state-of-the-art}
In the third set of experiments, we used the best method from previous section, i.e. the one based on the GMC and compared it to two state-of-the-art methods: 1) the ship wake detection method proposed by Graziano et. al. \cite{graziano1}, ii) the log-regularised Hough transform based method (Log-Hough) \cite{aggarwal2006line}. The results are presented in Table \ref{tab:comparisonMethodsAll} and \ref{tab:comparisonMethodsEach}.
\begin{figure}[ht!]
\centering
\includegraphics[width=.65\linewidth]{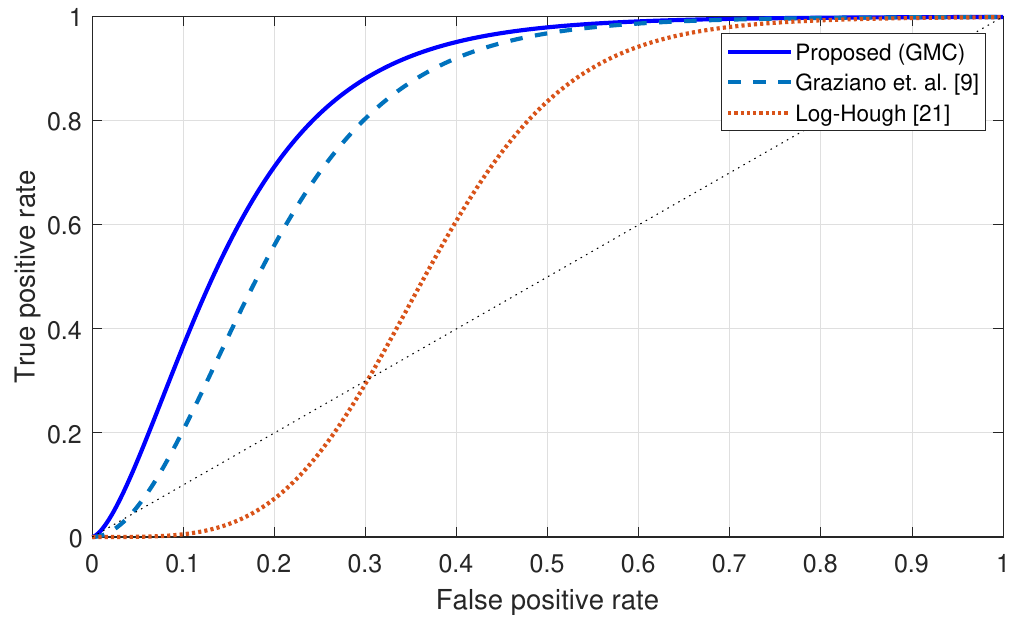}
\caption{Performance comparison in terms of a receiver operating characteristics (ROC) curve.}
\label{fig:rocMethods}
\end{figure}

The performance of all three methods over all data sets are presented in Table \ref{tab:comparisonMethodsAll} in terms of the same metrics discussed in previous section. The proposed method outperforms the reference methods by 10\% in terms of accuracy and to various degree about the other performance metrics presented. Specifically, the TP value is higher than for the other methods, whereas the TN value is slightly lower than Graziano et. al. \cite{graziano1}.

\begin{table*}[htbp]
  \centering
  \caption{Detection performance comparison of methods over all data sets.}
    \resizebox{.84\linewidth}{!}{\begin{tabular}{ccccccccccccccccccccccccccccccccccccccccc}
    \hline
          &       &       &       & \multicolumn{4}{c}{TP}        & \multicolumn{4}{c}{TN}        & \multicolumn{4}{c}{FP}        & \multicolumn{4}{c}{FN}        & \multicolumn{4}{c}{Sensitivity} & \multicolumn{4}{c}{Specificity} & \multicolumn{4}{|c}{\% Accuracy}  & \multicolumn{3}{c}{$F_1$} & \multicolumn{3}{c}{LR+} & \multicolumn{3}{c}{Youden's $J$} \\
          \hline
    \multicolumn{4}{c}{Proposed (GMC)} & \multicolumn{4}{c}{49.29\%}   & \multicolumn{4}{c}{30.71\%}   & \multicolumn{4}{c}{17.86\%}   & \multicolumn{4}{c}{2.14\%}    & \multicolumn{4}{c}{95.83\%}   & \multicolumn{4}{c}{63.24\%}   & \multicolumn{4}{|c}{\textbf{80.00\%}} & \multicolumn{3}{c}{\textbf{0.83}} & \multicolumn{3}{c}{\textbf{2.61}} & \multicolumn{3}{c}{\textbf{0.59}} \\
    \multicolumn{4}{c}{Graziano et. al. \cite{graziano1}}  & \multicolumn{4}{c}{33.57\%}   & \multicolumn{4}{c}{35.71\%}   & \multicolumn{4}{c}{26.43\%}   & \multicolumn{4}{c}{4.29\%}    & \multicolumn{4}{c}{88.68\%}   & \multicolumn{4}{c}{57.47\%}   & \multicolumn{4}{|c}{69.29\%}   & \multicolumn{3}{c}{0.69} & \multicolumn{3}{c}{2.09} & \multicolumn{3}{c}{0.46} \\
    \multicolumn{4}{c}{Log-Hough \cite{aggarwal2006line}}  & \multicolumn{4}{c}{38.57\%}   & \multicolumn{4}{c}{25.71\%}   & \multicolumn{4}{c}{33.57\%}   & \multicolumn{4}{c}{2.14\%}    & \multicolumn{4}{c}{94.74\%}   & \multicolumn{4}{c}{43.37\%}   & \multicolumn{4}{|c}{64.29\%}   & \multicolumn{3}{c}{0.68} & \multicolumn{3}{c}{1.67} & \multicolumn{3}{c}{0.38} \\
    \hline
    \end{tabular}}%
  \label{tab:comparisonMethodsAll}%
\end{table*}%

\begin{table*}[htbp]
  \centering
  \caption{Detection performance comparison of methods for each data sets separately.}
    \resizebox{.84\linewidth}{!}{\begin{tabular}{ccccccccccccccccccccccccccccccccccccccccc}
    \hline
          &       &       &       & \multicolumn{37}{c}{TerraSAR-X} \\
          \hline
          &       &       &       & \multicolumn{4}{c}{TP}        & \multicolumn{4}{c}{TN}        & \multicolumn{4}{c}{FP}        & \multicolumn{4}{c}{FN}        & \multicolumn{4}{c}{Sensitivity} & \multicolumn{4}{c}{Specificity} & \multicolumn{4}{|c}{\% Accuracy}  & \multicolumn{3}{c}{$F_1$} & \multicolumn{3}{c}{LR+} & \multicolumn{3}{c}{Youden's $J$} \\
          \hline
    \multicolumn{4}{c}{Proposed (GMC)}  & \multicolumn{4}{c}{61.67\%}   & \multicolumn{4}{c}{21.67\%}   & \multicolumn{4}{c}{13.33\%}   & \multicolumn{4}{c}{3.33\%}    & \multicolumn{4}{c}{94.87\%}   & \multicolumn{4}{c}{61.90\%}   & \multicolumn{4}{|c}{\textbf{83.33\%}} & \multicolumn{3}{c}{\textbf{0.88}} & \multicolumn{3}{c}{\textbf{2.49}} & \multicolumn{3}{c}{\textbf{0.57}} \\
    \multicolumn{4}{c}{Graziano et. al. \cite{graziano1}}  & \multicolumn{4}{c}{35.00\%}   & \multicolumn{4}{c}{23.33\%}   & \multicolumn{4}{c}{33.33\%}   & \multicolumn{4}{c}{8.33\%}    & \multicolumn{4}{c}{80.77\%}   & \multicolumn{4}{c}{41.18\%}   & \multicolumn{4}{|c}{58.33\%}   & \multicolumn{3}{c}{0.63} & \multicolumn{3}{c}{1.37} & \multicolumn{3}{c}{0.22} \\
    \multicolumn{4}{c}{Log-Hough \cite{aggarwal2006line}}  & \multicolumn{4}{c}{48.33\%}   & \multicolumn{4}{c}{15.00\%}   & \multicolumn{4}{c}{33.33\%}   & \multicolumn{4}{c}{3.33\%}    & \multicolumn{4}{c}{93.55\%}   & \multicolumn{4}{c}{31.03\%}   & \multicolumn{4}{|c}{63.33\%}   & \multicolumn{3}{c}{0.73} & \multicolumn{3}{c}{1.36} & \multicolumn{3}{c}{0.25} \\
    \hline
          \hline
          \multicolumn{4}{c}{}          & \multicolumn{37}{c}{ALOS2} \\
          \hline
          &       &       &       & \multicolumn{4}{c}{TP}        & \multicolumn{4}{c}{TN}        & \multicolumn{4}{c}{FP}        & \multicolumn{4}{c}{FN}        & \multicolumn{4}{c}{Sensitivity} & \multicolumn{4}{c}{Specificity} & \multicolumn{4}{|c}{\% Accuracy}  & \multicolumn{3}{c}{$F_1$} & \multicolumn{3}{c}{LR+} & \multicolumn{3}{c}{Youden's $J$} \\
          \hline
    \multicolumn{4}{c}{Proposed (GMC)}  & \multicolumn{4}{c}{40.00\%}   & \multicolumn{4}{c}{35.00\%}   & \multicolumn{4}{c}{25.00\%}   & \multicolumn{4}{c}{0.00\%}    & \multicolumn{4}{c}{100.00\%}  & \multicolumn{4}{c|}{58.33\%}  & \multicolumn{4}{c}{75.00\%}   & \multicolumn{3}{c}{0.76} & \multicolumn{3}{c}{2.40} & \multicolumn{3}{c}{0.58} \\
    \multicolumn{4}{c}{Graziano et. al. \cite{graziano1}}  & \multicolumn{4}{c}{30.00\%}   & \multicolumn{4}{c}{40.00\%}   & \multicolumn{4}{c}{30.00\%}   & \multicolumn{4}{c}{0.00\%}    & \multicolumn{4}{c}{100.00\%}  & \multicolumn{4}{c|}{57.14\%}  & \multicolumn{4}{c}{70.00\%}   & \multicolumn{3}{c}{0.67} & \multicolumn{3}{c}{2.33} & \multicolumn{3}{c}{0.57} \\
    \multicolumn{4}{c}{Log-Hough \cite{aggarwal2006line}}  & \multicolumn{4}{c}{45.00\%}   & \multicolumn{4}{c}{35.00\%}   & \multicolumn{4}{c}{20.00\%}   & \multicolumn{4}{c}{0.00\%}    & \multicolumn{4}{c}{100.00\%}  & \multicolumn{4}{c|}{63.64\%}  & \multicolumn{4}{c}{\textbf{80.00\%}} & \multicolumn{3}{c}{\textbf{0.82}} & \multicolumn{3}{c}{\textbf{2.75}} & \multicolumn{3}{c}{\textbf{0.64}} \\
    \hline
    \hline
          &       &       &       & \multicolumn{37}{c}{COSMO-SkyMed} \\
          \hline
          &       &       &       & \multicolumn{4}{c}{TP}        & \multicolumn{4}{c}{TN}        & \multicolumn{4}{c}{FP}        & \multicolumn{4}{c}{FN}        & \multicolumn{4}{c}{Sensitivity} & \multicolumn{4}{c}{Specificity} & \multicolumn{4}{|c}{\% Accuracy}  & \multicolumn{3}{c}{$F_1$} & \multicolumn{3}{c}{LR+} & \multicolumn{3}{c}{Youden's $J$} \\
          \hline
    \multicolumn{4}{c}{Proposed (GMC)}  & \multicolumn{4}{c}{45.00\%}   & \multicolumn{4}{c}{40.00\%}   & \multicolumn{4}{c}{10.00\%}   & \multicolumn{4}{c}{5.00\%}    & \multicolumn{4}{c}{90.00\%}   & \multicolumn{4}{c|}{80.00\%}  & \multicolumn{4}{c}{\textbf{85.00\%}} & \multicolumn{3}{c}{\textbf{0.86}} & \multicolumn{3}{c}{\textbf{4.50}} & \multicolumn{3}{c}{\textbf{0.70}} \\
    \multicolumn{4}{c}{Graziano et. al. \cite{graziano1}}  & \multicolumn{4}{c}{40.00\%}   & \multicolumn{4}{c}{40.00\%}   & \multicolumn{4}{c}{15.00\%}   & \multicolumn{4}{c}{5.00\%}    & \multicolumn{4}{c}{88.89\%}   & \multicolumn{4}{c|}{72.73\%}  & \multicolumn{4}{c}{80.00\%}   & \multicolumn{3}{c}{0.80} & \multicolumn{3}{c}{3.26} & \multicolumn{3}{c}{0.62} \\
    \multicolumn{4}{c}{Log-Hough \cite{aggarwal2006line}}  & \multicolumn{4}{c}{35.00\%}   & \multicolumn{4}{c}{30.00\%}   & \multicolumn{4}{c}{30.00\%}   & \multicolumn{4}{c}{5.00\%}    & \multicolumn{4}{c}{87.50\%}   & \multicolumn{4}{c|}{50.00\%}  & \multicolumn{4}{c}{65.00\%}   & \multicolumn{3}{c}{0.67} & \multicolumn{3}{c}{1.75} & \multicolumn{3}{c}{0.38} \\
    \hline
    \hline
          &       &       &       & \multicolumn{37}{c}{Sentinel-1} \\
          \hline
          &       &       &       & \multicolumn{4}{c}{TP}        & \multicolumn{4}{c}{TN}        & \multicolumn{4}{c}{FP}        & \multicolumn{4}{c}{FN}        & \multicolumn{4}{c}{Sensitivity} & \multicolumn{4}{c}{Specificity} & \multicolumn{4}{|c}{\% Accuracy}  & \multicolumn{3}{c}{$F_1$} & \multicolumn{3}{c}{LR+} & \multicolumn{3}{c}{Youden's $J$} \\
          \hline
    \multicolumn{4}{c}{Proposed (GMC)}  & \multicolumn{4}{c}{37.50\%}   & \multicolumn{4}{c}{37.50\%}   & \multicolumn{4}{c}{25.00\%}   & \multicolumn{4}{c}{0.00\%}    & \multicolumn{4}{c}{100.00\%}  & \multicolumn{4}{c|}{60.00\%}  & \multicolumn{4}{c}{75.00\%}   & \multicolumn{3}{c}{\textbf{0.75}} & \multicolumn{3}{c}{2.50} & \multicolumn{3}{c}{0.60} \\
    \multicolumn{4}{c}{Graziano et. al. \cite{graziano1}}  & \multicolumn{4}{c}{30.00\%}  & \multicolumn{4}{c}{50.00\%}   & \multicolumn{4}{c}{20.00\%}   & \multicolumn{4}{c}{0.00\%}    & \multicolumn{4}{c}{100.00\%}  & \multicolumn{4}{c|}{71.43\%}  & \multicolumn{4}{c}{\textbf{80.00\%}} & \multicolumn{3}{c}{\textbf{0.75}} & \multicolumn{3}{c}{\textbf{3.50}} & \multicolumn{3}{c}{\textbf{0.71}} \\
    \multicolumn{4}{c}{Log-Hough \cite{aggarwal2006line}}  & \multicolumn{4}{c}{22.50\%}   & \multicolumn{4}{c}{35.00\%}   & \multicolumn{4}{c}{42.50\%}   & \multicolumn{4}{c}{0.00\%}    & \multicolumn{4}{c}{100.00\%}  & \multicolumn{4}{c|}{45.16\%}  & \multicolumn{4}{c}{57.50\%}   & \multicolumn{3}{c}{0.51} & \multicolumn{3}{c}{1.82} & \multicolumn{3}{c}{0.45} \\
    \hline
    \end{tabular}}%
  \label{tab:comparisonMethodsEach}%
\end{table*}%

When examining the values in Table \ref{tab:comparisonMethodsEach} for each data source, the proposed method achieves at least 75\% detection performance for each satellite platforms. However, for Sentinel-1, the method of Graziano et. al achieves the best detection results. The lower performance of the proposed method for Sentinel-1 images can be explained in several ways. First, the resolution of Sentinel-1 images are less than that of the other platforms (i.e. 10 meters). We believe that the low resolution affects the performance of the proposed method since visibility of the wakes is reduced in low resolution. Besides, as high frequency (smaller wavelength) SAR sensors determine more surface scattering from rough surfaces, images in L- (ALOS2) and C- (Sentinel-2) bands include less sea surface details than X band (TerraSAR-X and COSMO-SkyMed). This directly influenced the performance of the proposed method and determined the most accurate results to be obtained for X band images.

\begin{figure*}[htbp]
\centering
\subfigure[Ship centered tile]{
\includegraphics[width=.23\linewidth]{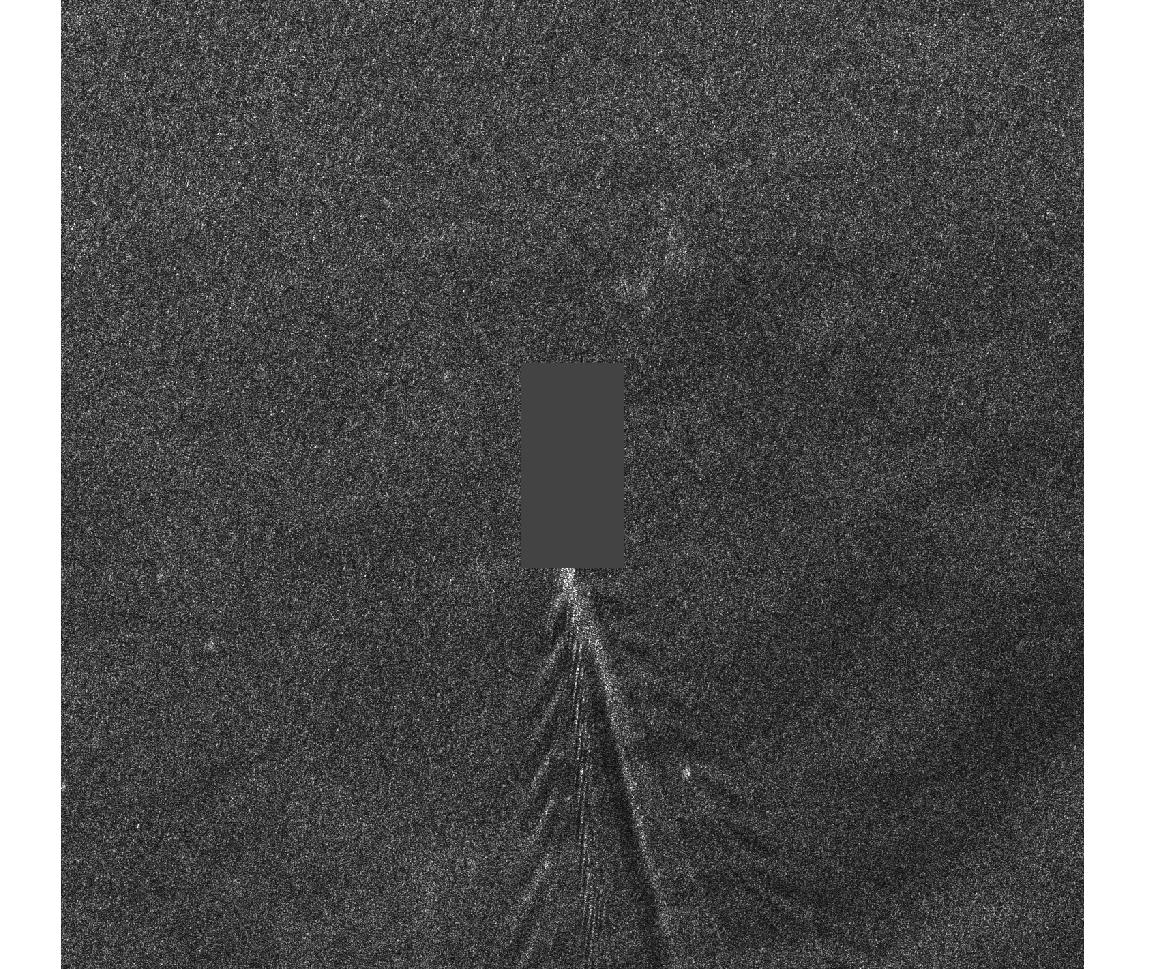}}
\centering
\subfigure[Proposed (GMC)]{
\includegraphics[width=.23\linewidth]{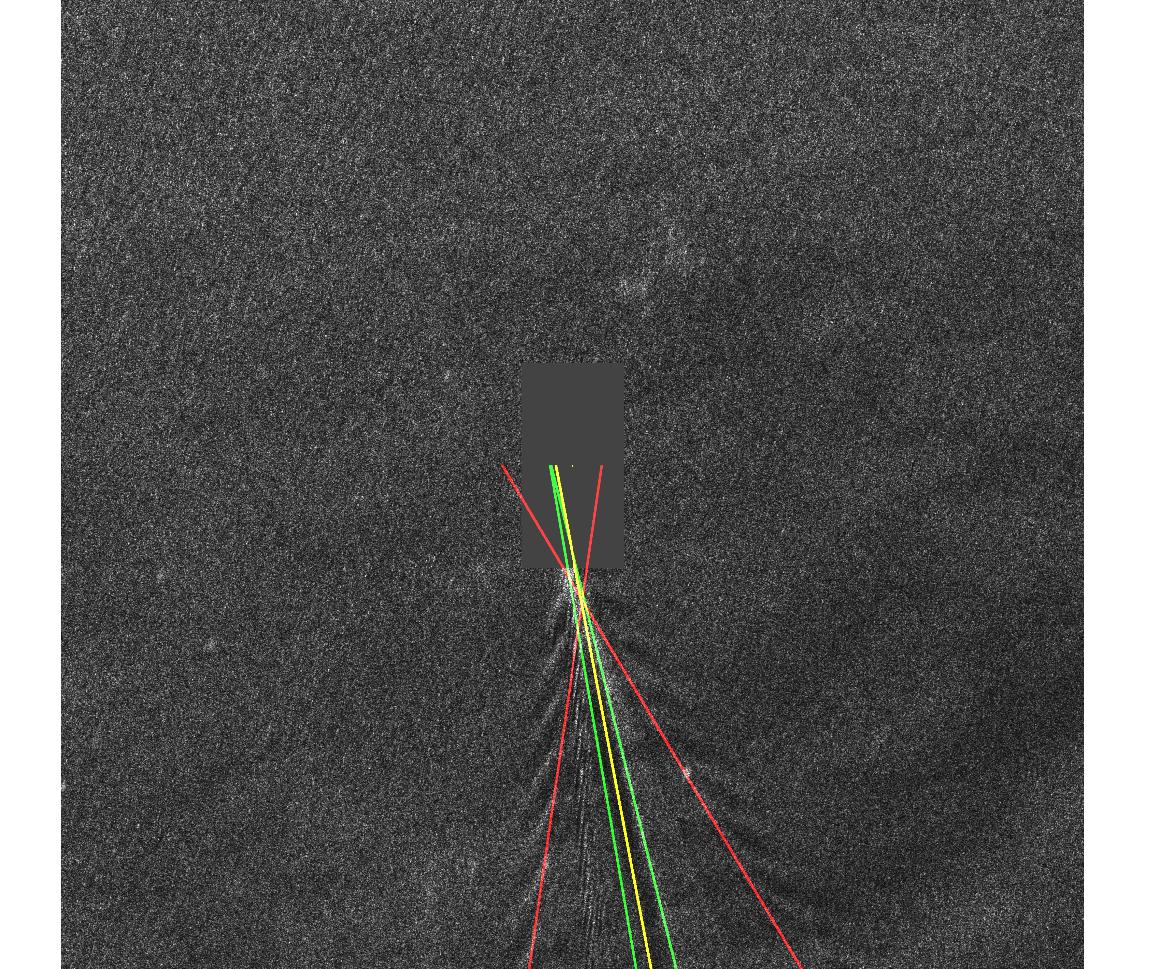}}
\centering
\subfigure[Graziano et. al. \cite{graziano1}]{%
\includegraphics[width=.23\linewidth]{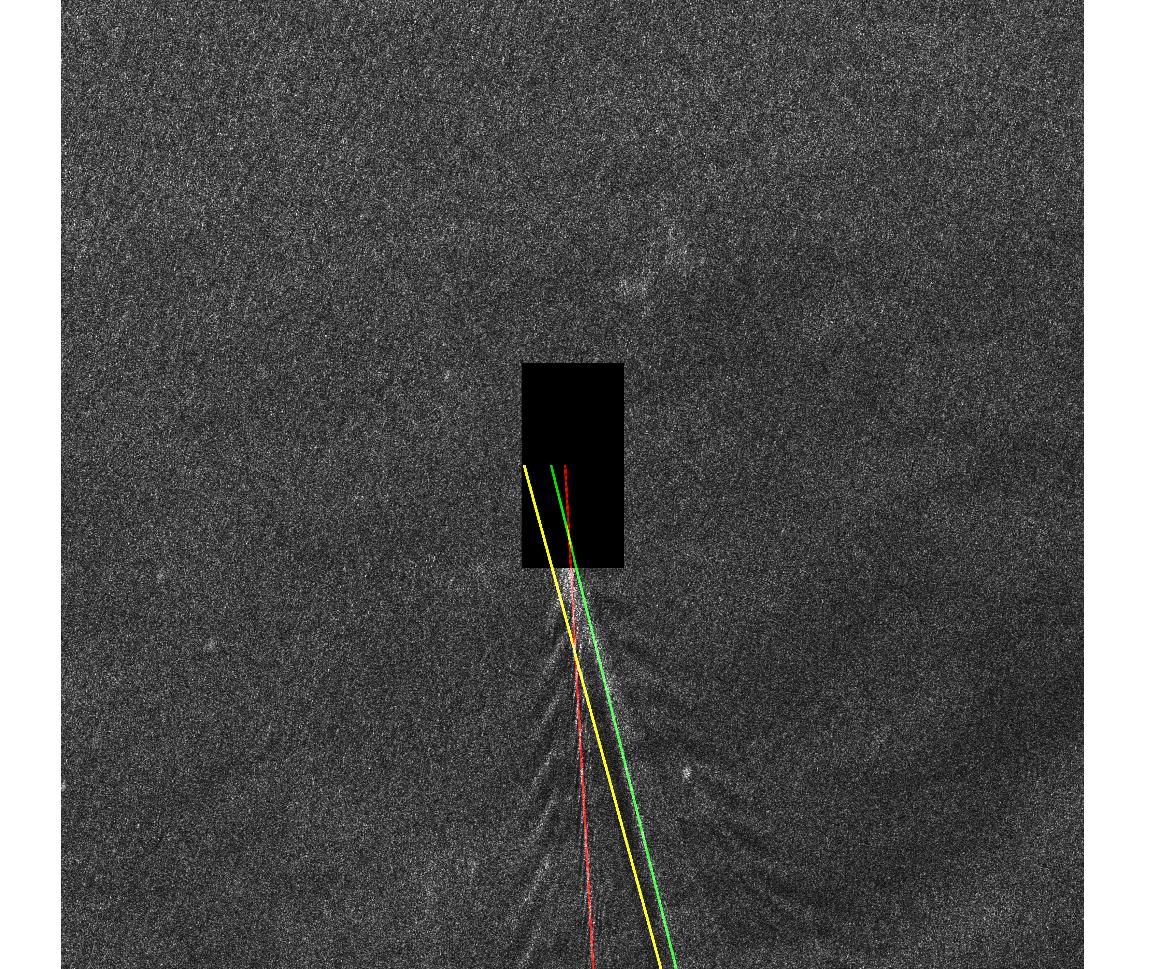}}
\centering
\subfigure[Log-Hough \cite{aggarwal2006line}]{%
\includegraphics[width=.23\linewidth]{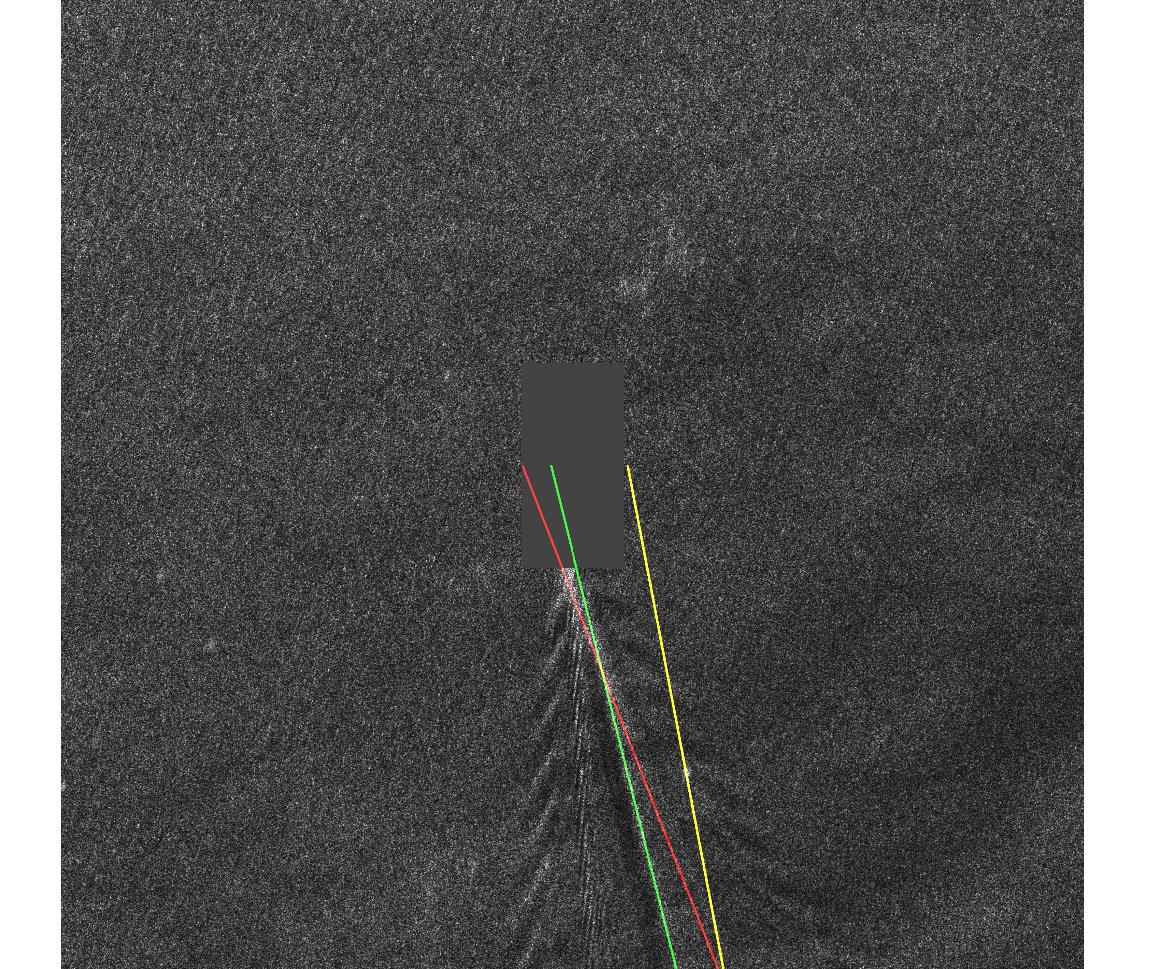}}
\centering
\subfigure[Ship centered tile]{
\includegraphics[width=.23\linewidth]{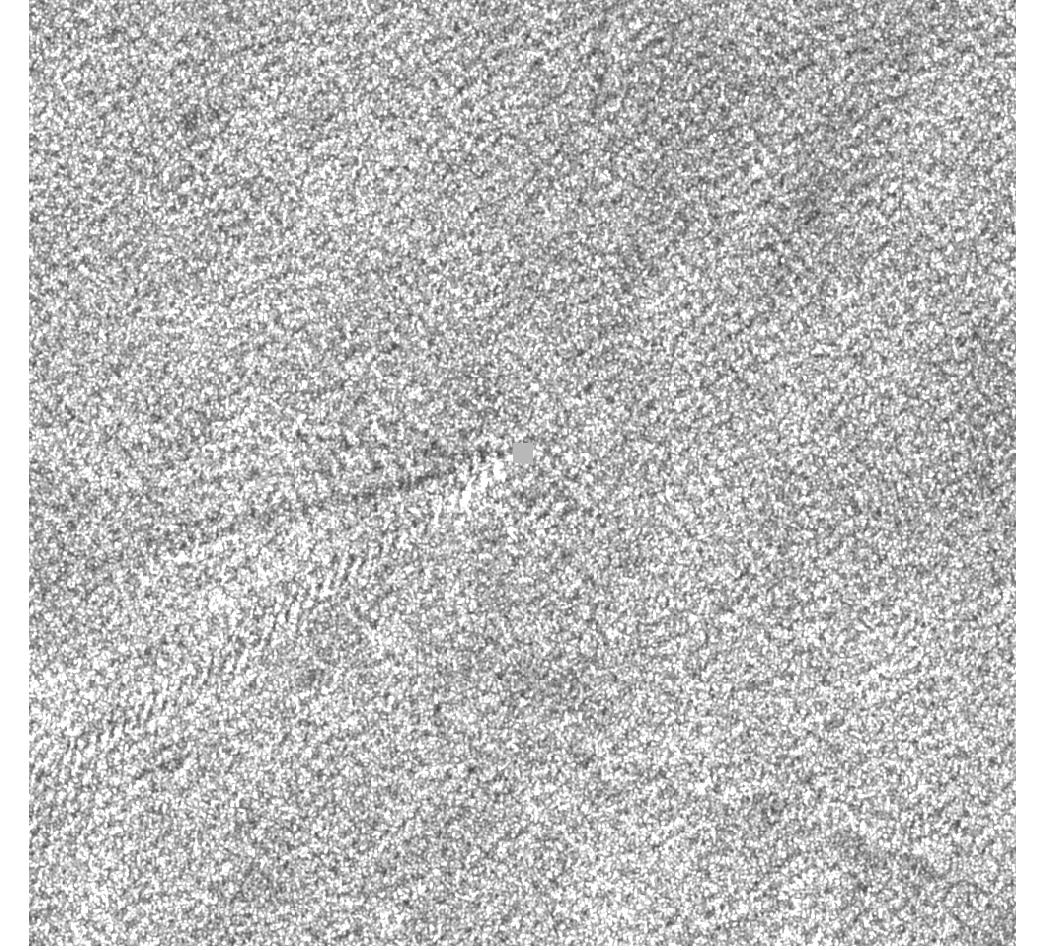}}
\centering
\subfigure[Proposed (GMC)]{
\includegraphics[width=.23\linewidth]{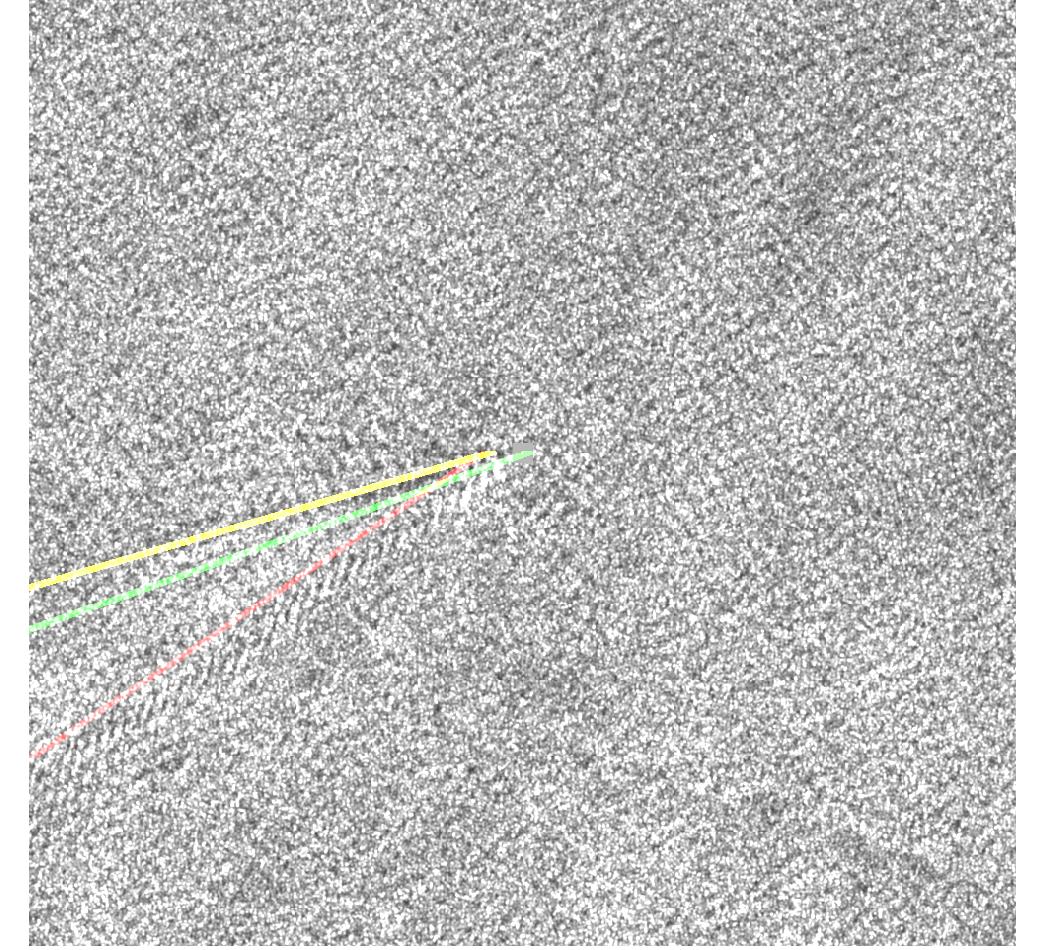}}
\centering
\subfigure[Graziano et. al. \cite{graziano1}]{%
\includegraphics[width=.23\linewidth]{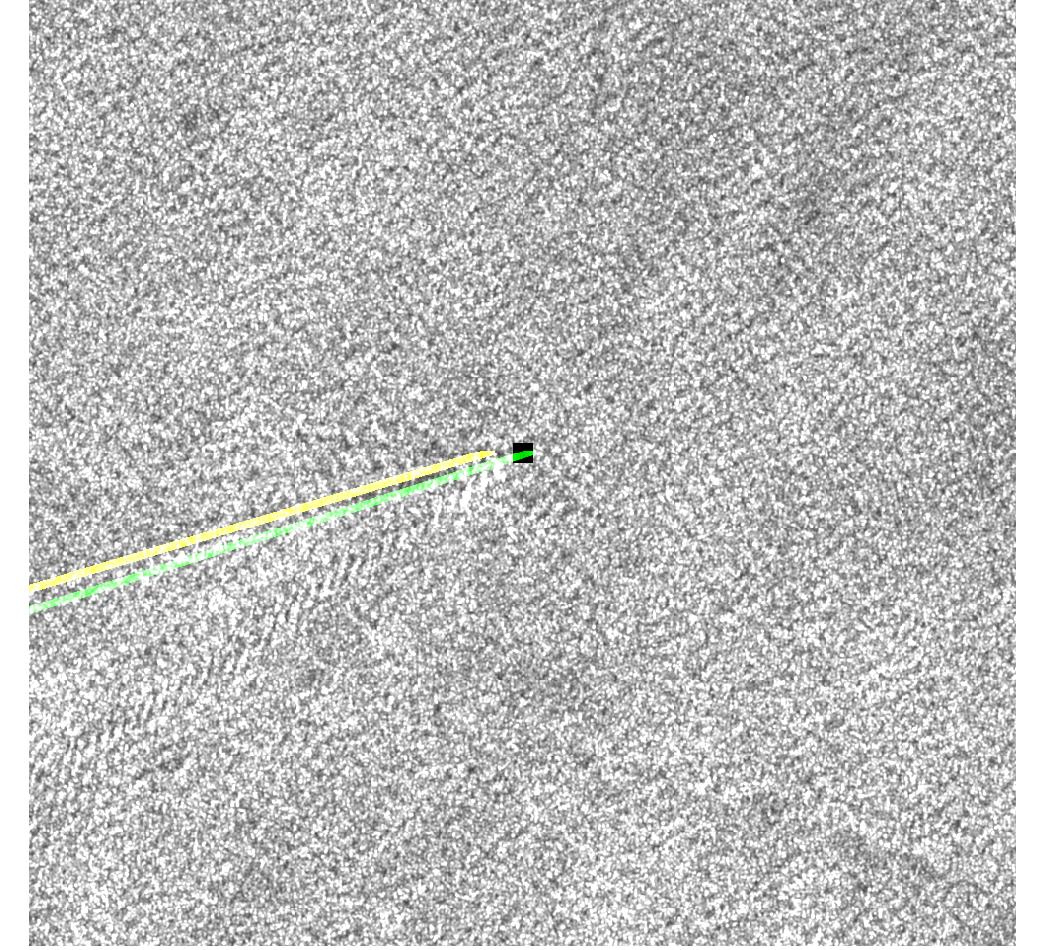}}
\centering
\subfigure[Log-Hough \cite{aggarwal2006line}]{%
\includegraphics[width=.23\linewidth]{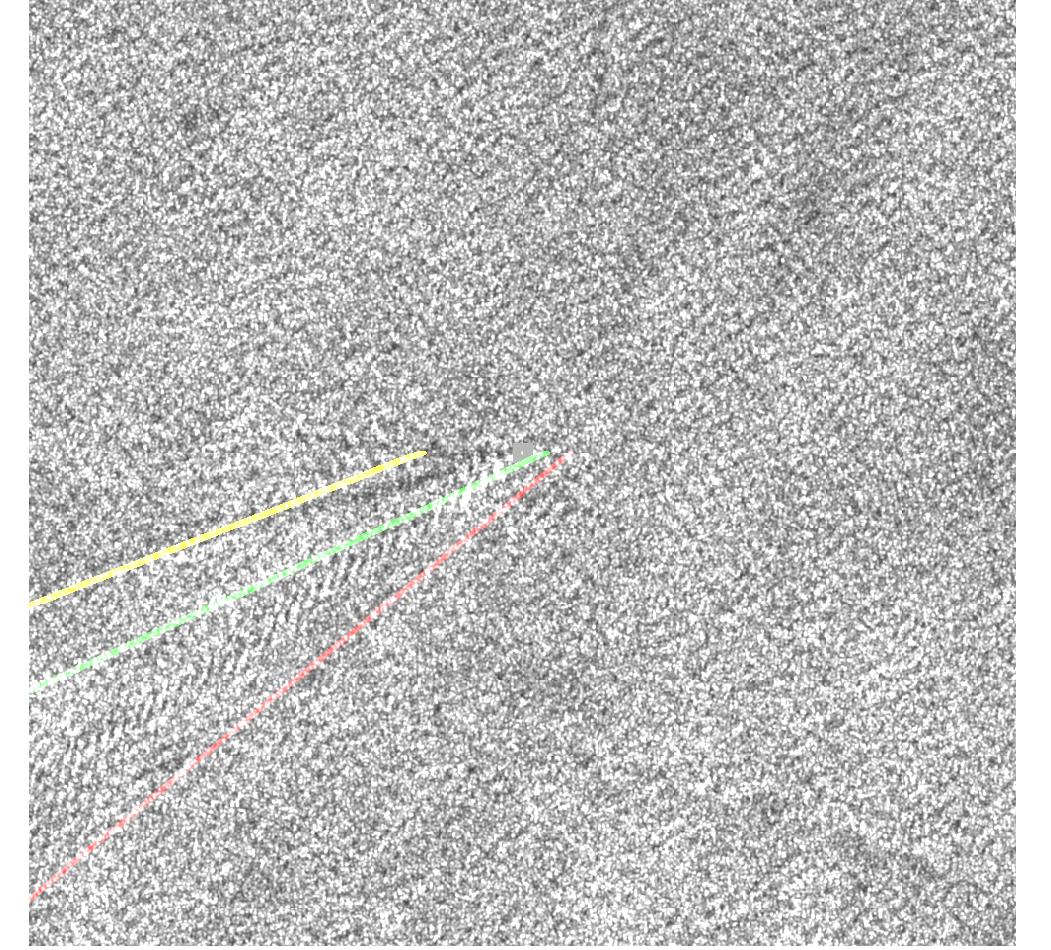}}
\centering
\subfigure[Ship centered tile]{
\includegraphics[width=.23\linewidth]{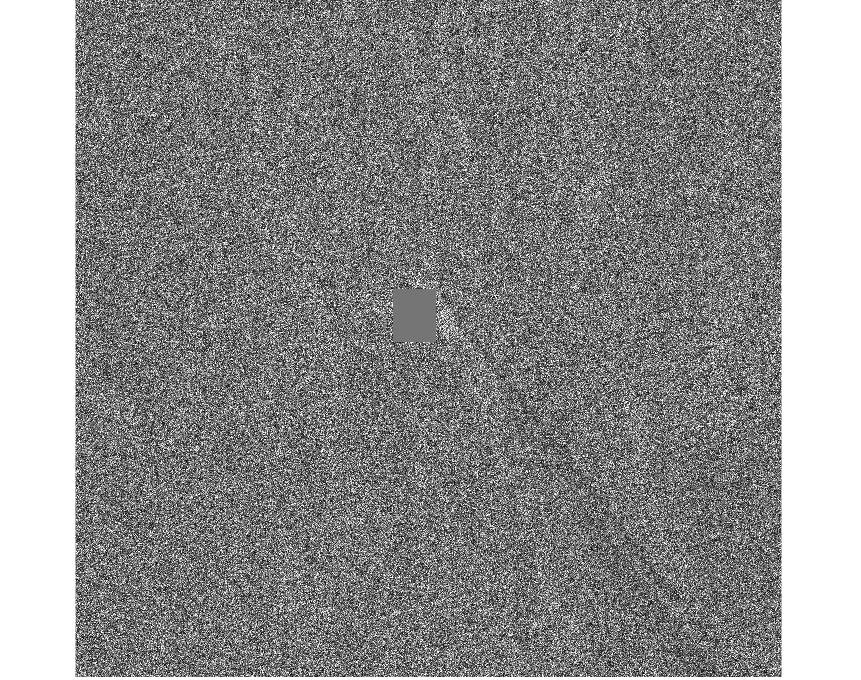}}
\centering
\subfigure[Proposed (GMC)]{
\includegraphics[width=.23\linewidth]{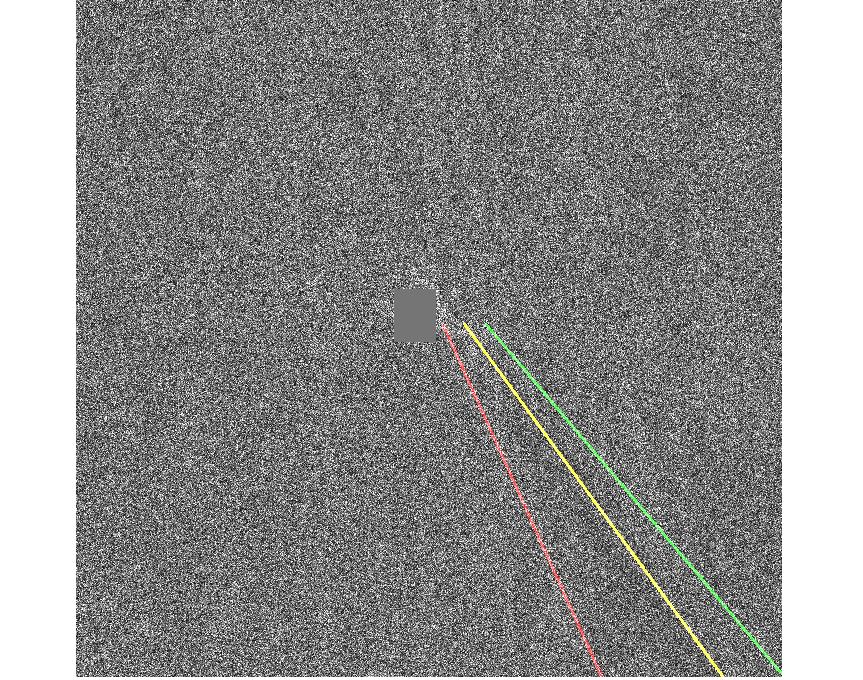}}
\centering
\subfigure[Graziano et. al. \cite{graziano1}]{%
\includegraphics[width=.23\linewidth]{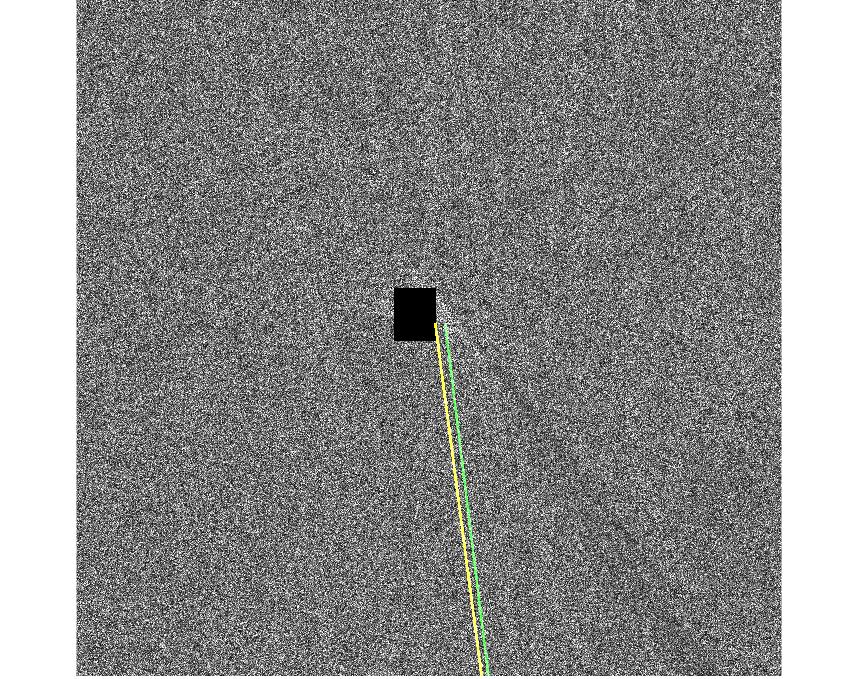}}
\centering
\subfigure[Log-Hough \cite{aggarwal2006line}]{%
\includegraphics[width=.23\linewidth]{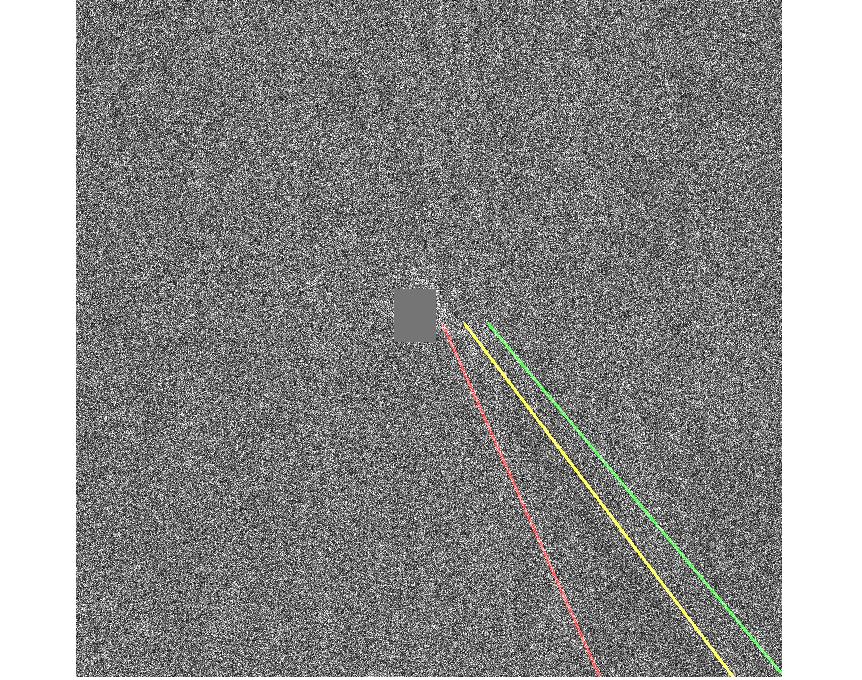}}
\centering
\subfigure[Ship centered tile]{
\includegraphics[width=.23\linewidth]{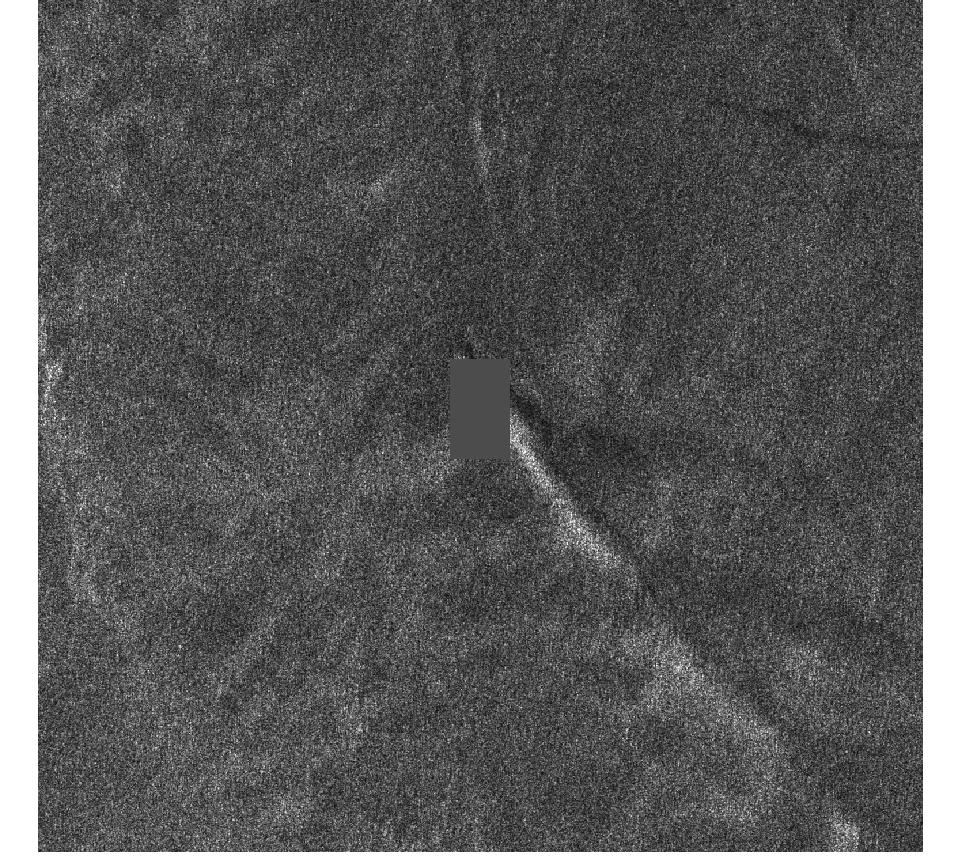}}
\centering
\subfigure[Proposed (GMC)]{
\includegraphics[width=.23\linewidth]{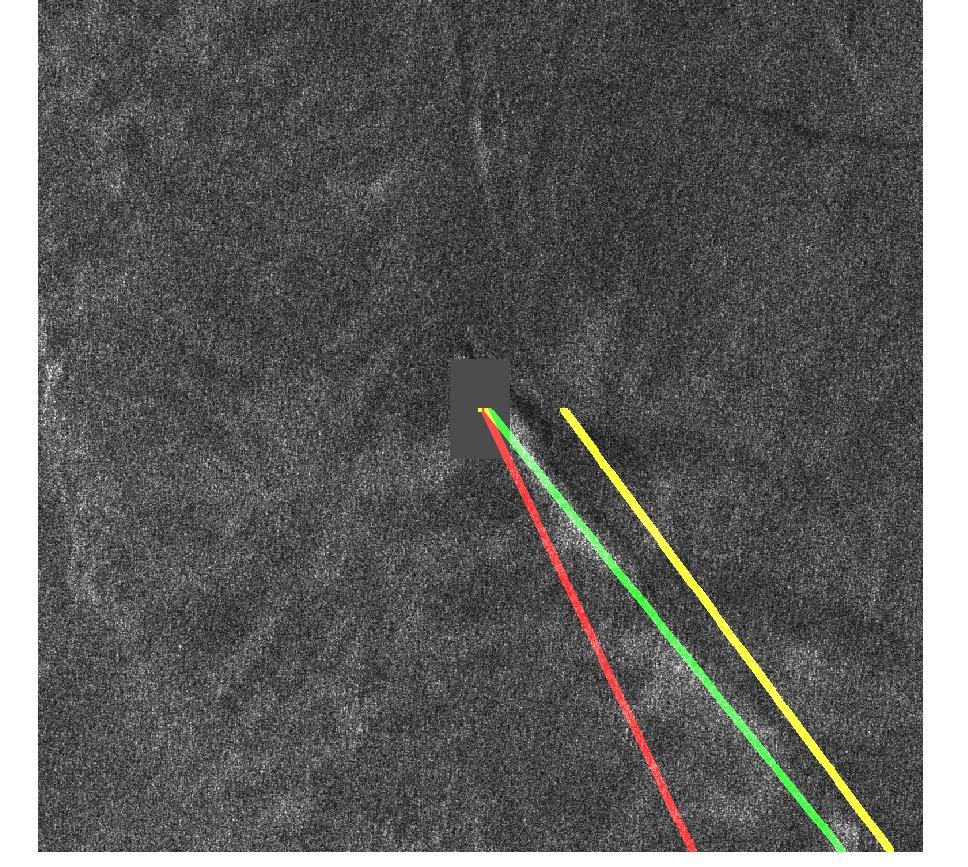}}
\centering
\subfigure[Graziano et. al. \cite{graziano1}]{%
\includegraphics[width=.23\linewidth]{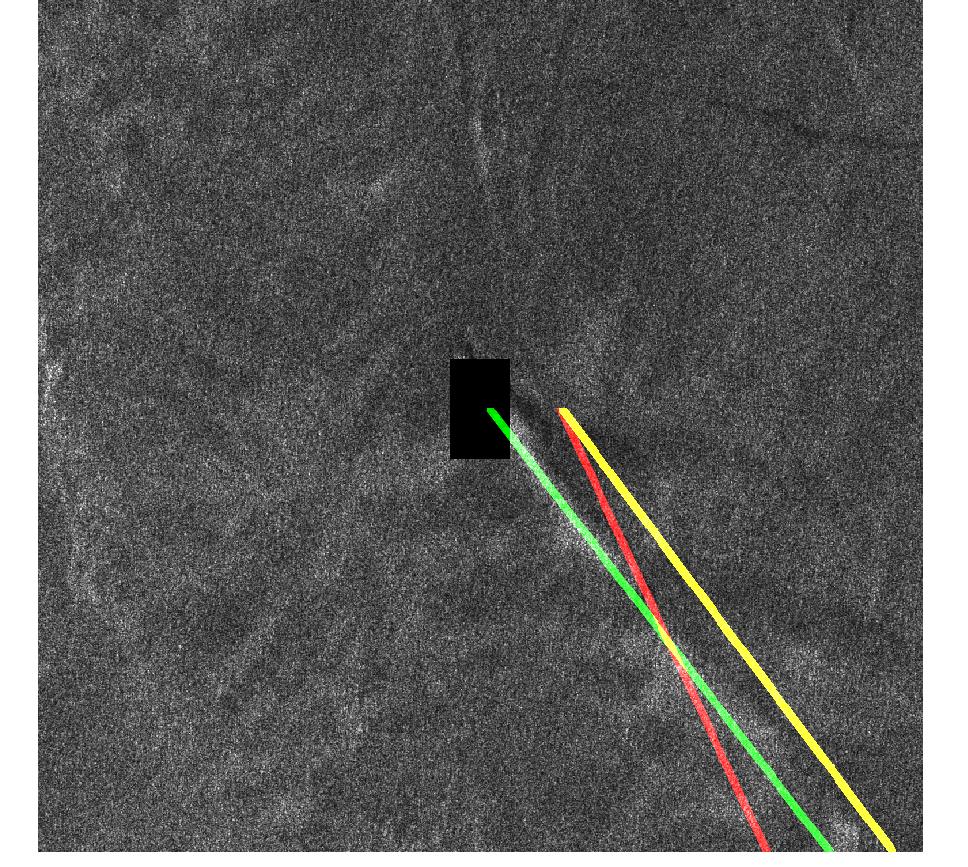}}
\centering
\subfigure[Log-Hough \cite{aggarwal2006line}]{%
\includegraphics[width=.23\linewidth]{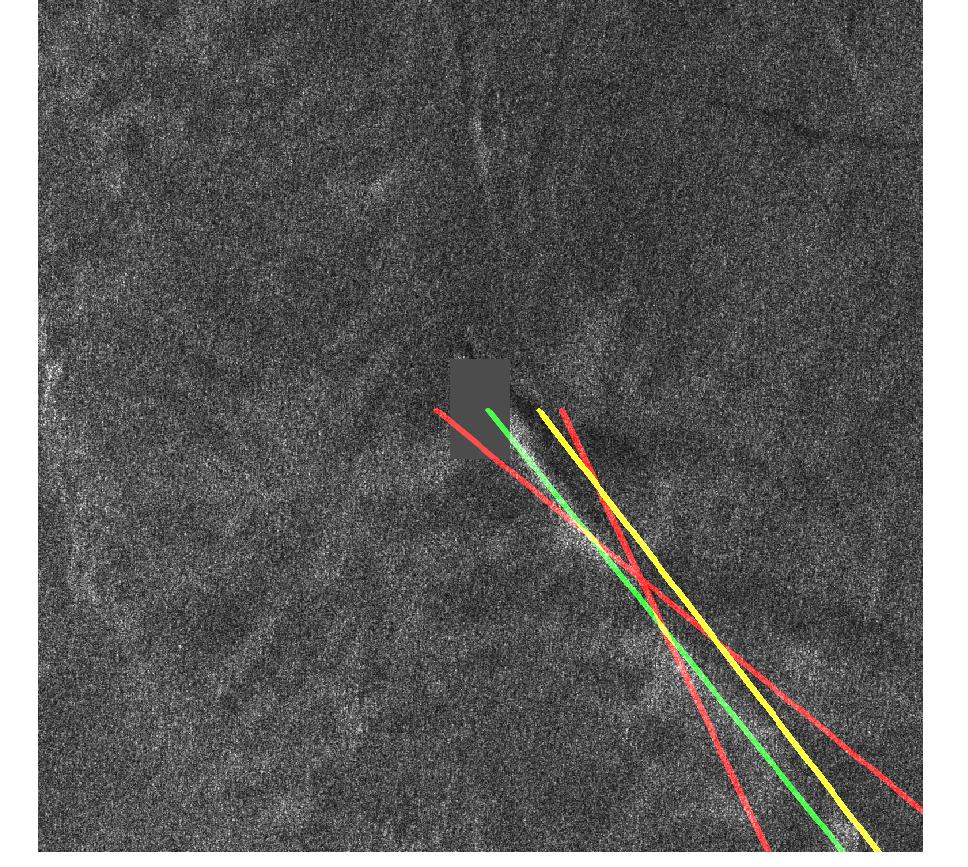}}
\centering
\subfigure[Ship centered tile]{
\includegraphics[width=.23\linewidth]{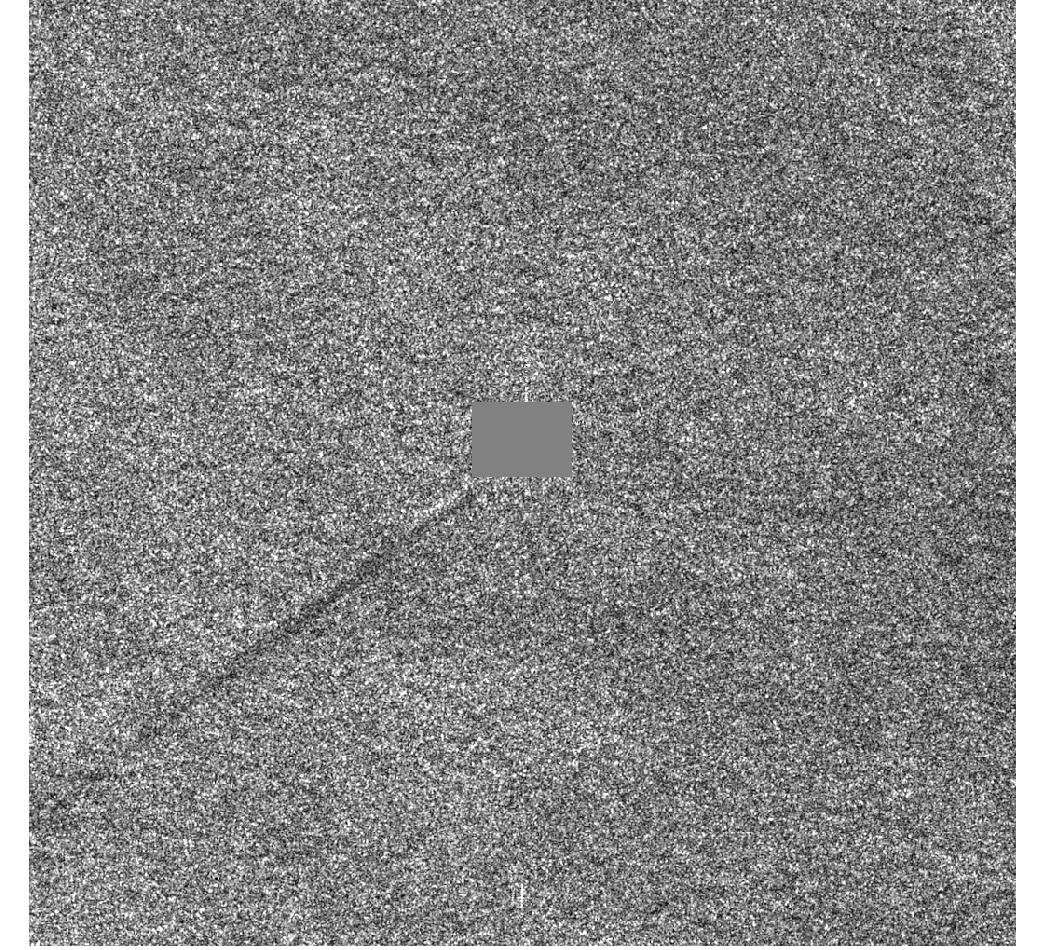}}
\centering
\subfigure[Proposed (GMC)]{
\includegraphics[width=.23\linewidth]{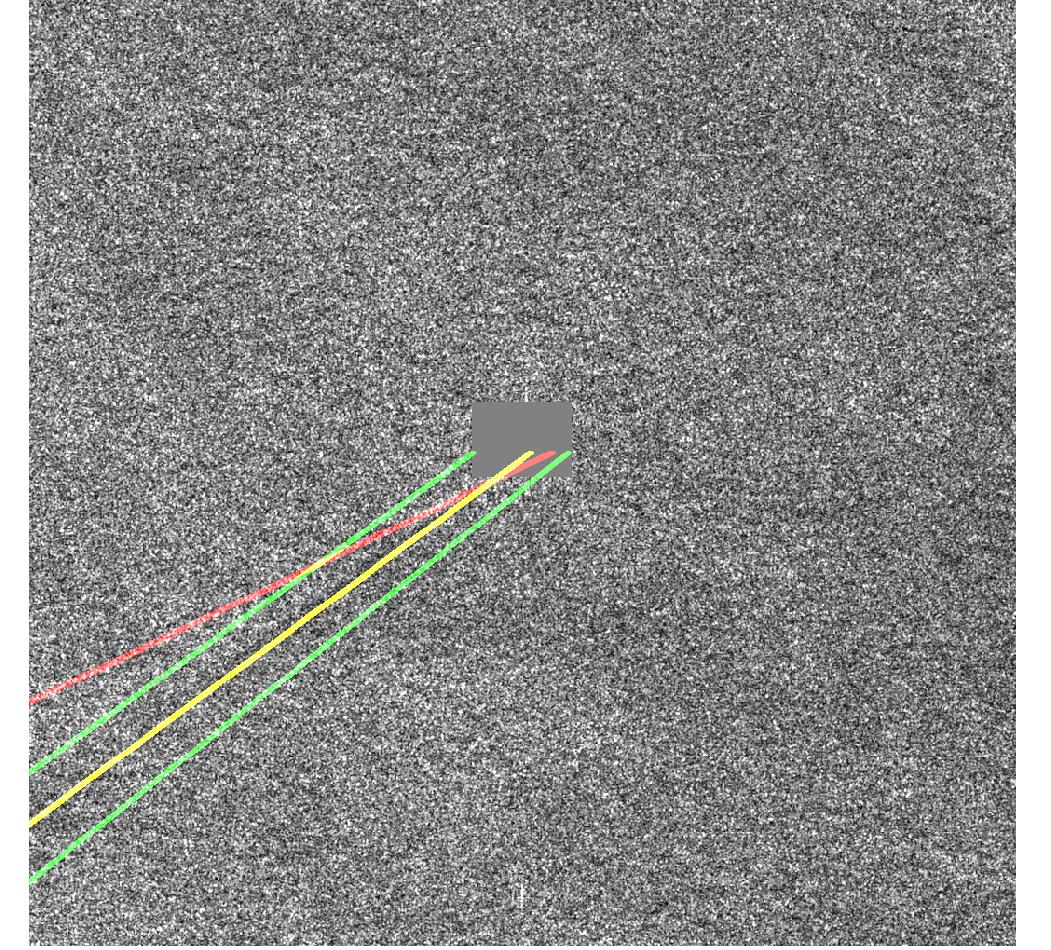}}
\centering
\subfigure[Graziano et. al. \cite{graziano1}]{%
\includegraphics[width=.23\linewidth]{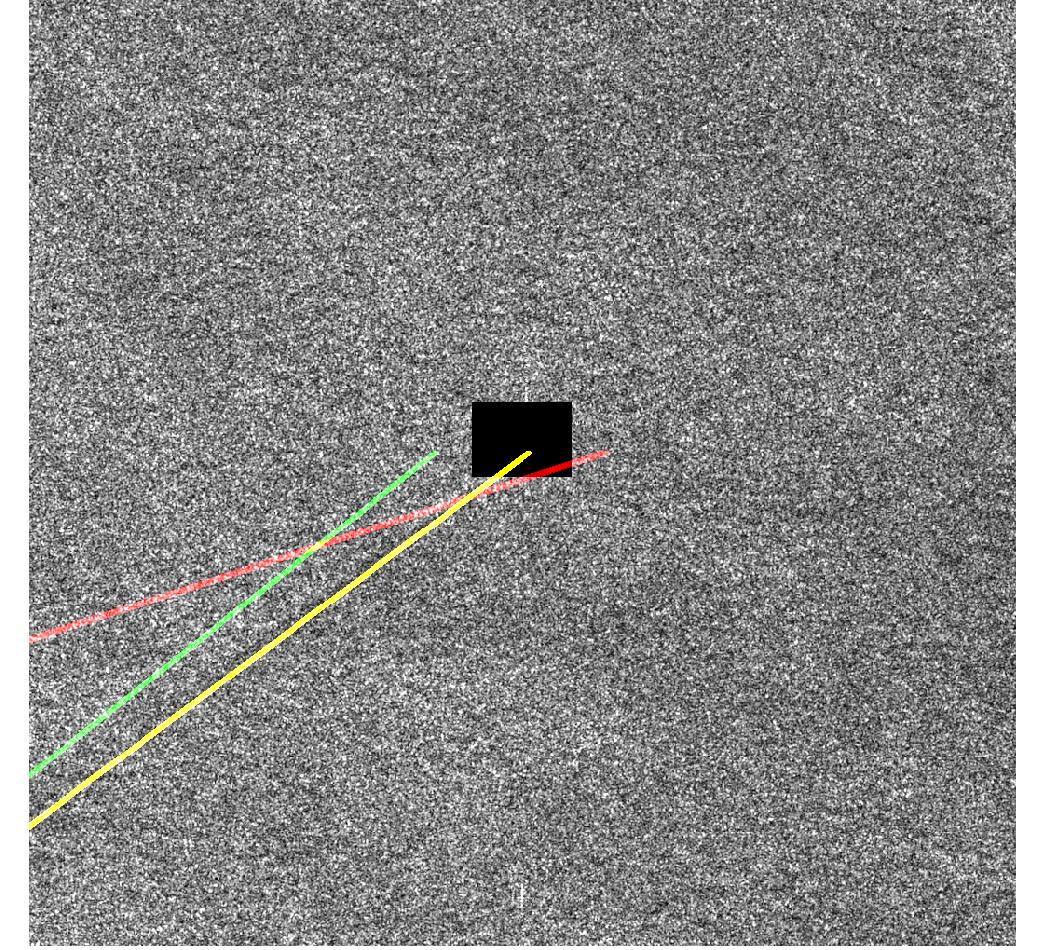}}
\centering
\subfigure[Log-Hough \cite{aggarwal2006line}]{%
\includegraphics[width=.23\linewidth]{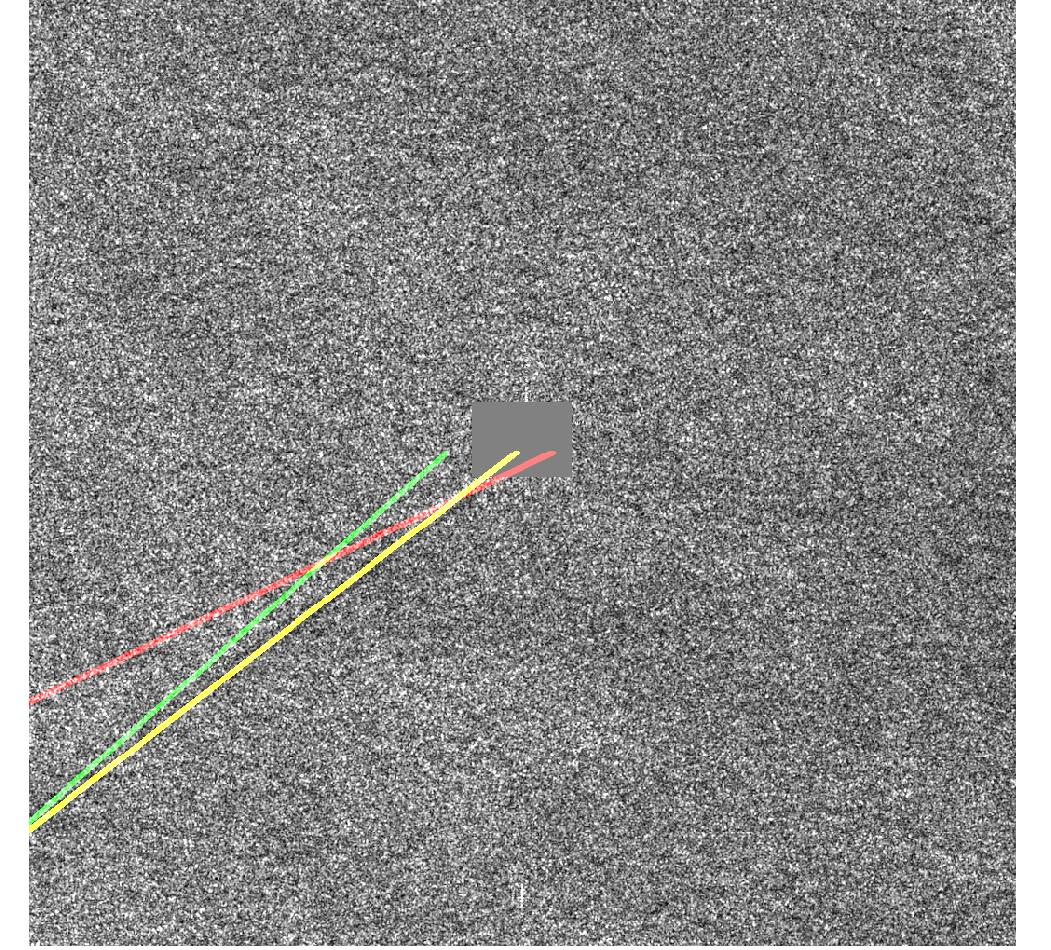}}
\caption{Visual ship wake detection results. From top to bottom, each line represents wakes 1.1, 3.3, 4.1, 6.1 and 8.2., respectively.}
\label{fig:methodsDetection4}
\end{figure*}

When examining the TN values in Table \ref{tab:comparisonMethodsEach} for Sentinel-1 images, the proposed method achieves 37.5\% whereas the method of Graziano et. al. 50\%, even though the TP value of the proposed method is the highest. Enhancing the image in Radon space using the proposed methodology increased the detectability of the visible wakes (TP values) in the images. However, in discarding the invisible wakes (TN values), Graziano's method is generally better than the proposed method whereas it falls short in detecting the visible wakes. The values of FP and FN are also related to the Eldhuset's \cite{eldhuset1996automatic} performance metrics that quantify false and lost wakes. In particular, FP and FN can be defined as the percentage of false and lost wakes, respectively. Examining FN and FP values, we can clearly see that the proposed method is substantially better than the reference methods in terms of FP results (false wakes), which is obvious especially in TerraSAR-X images where it results in 20\% higher wake detection accuracy. For all the methods, FN values (the number of lost wakes) are relatively small and do not directly correlate with performance. Example detection results for various images are presented in Figure \ref{fig:methodsDetection4}.

In Figure \ref{fig:rocMethods}, ROC curves for the methods tested in this section are presented. The superiority of the proposed method compared to existing ones is obvious. Graziano et. al. \cite{graziano1} is the closest to the proposed method in terms of ROC curve. The Log-Hough falls short compared to the proposed method and Graziano et. al, which is not surprising since the log-regularisation in \cite{aggarwal2006line} converges to an $L_p$ when $p$ tends to 0.

\section{Conclusions}\label{sec:conc}
In this paper, we proposed a novel automatic approach for ship wake detection in SAR images of the sea surface acquired by various satellite platforms including TerraSAR-X, ALOS2, COSMO-SkyMed and Sentinel-1. The proposed approach handles wake detection as an inverse problem to enhance information in Radon domain to promote linear features. The solution to the corresponding optimisation problem is obtained via a Bayesian formulation using a MAP estimator. The proposed method based on the GMC prior was first compared to various benchmark priors which are based on $L_1$, $L_p$, $TV$, and the nuclear norms. The GMC based method was then compared to two state-of-the-art methods including Graziano et. al. \cite{graziano1,graziano2} and Log-Hough \cite{aggarwal2006line}. The superiority of the GMC based method has been clearly demonstrated in both sets of simulations with at least 10\% accuracy gain over all data sets.

We conclude that enhancing sea SAR images in Radon space provides a more suitable platform for detecting the peaks/through compared to the direct approach in Graziano et. al. Nevertheless, merely solving an inverse problem is not sufficient to obtain the most accurate results, since the choice of the prior has a crucial effect on the results. Indeed, only two priors (GMC and TV) lead to higher accuracy than Graziano et. al.

Furthermore, the proposed approach differs from other approaches in the literature which require the use of some ad hoc thresholding for enhancing the image. The main tunable parameter in our method is the regularisation constant, $\lambda$. The need to adjust the scale parameter will be removed in future studies, which will consider a hierarchical Bayesian inference step. Investigating more complex sparsity enforcing priors in conjunction with non-convex optimisation algorithms is also one of our current endeavours.

\appendix\label{sec:app}
\section*{List of Abbreviations}
\vspace{0.1cm}
\begin{center}

    \resizebox{0.65\linewidth}{!}{\begin{tabular}{rcl}
    \hline
    Abbreviation && Description \\
    \hline
    SAR && Synthetic Aperture Radar\\
    MC && Minimax Concave\\
    GMC && Generalized Minimax Concave\\
    MAP && Maximum A-Posteriori\\
    COSMO-SkyMed && Constellation of small Satellites for the Mediterranean\\
    && basin Observation\\
    ALOS2 && Advanced Land Observing Satellite 2\\
    TV && Total Variation\\
    FB && Forward-Backward\\
    TwIST && Two-step Iterative Shrinkage/Thresholding\\
    MCMC && Markov Chain Monte Carlo\\
    p-MCMC && Proximal Markov Chain Monte Carlo\\
    GST && Generalised Soft Thresholding\\
    RAR && Real Asperture Radar\\
    NRCS && Normalised Radar Cross Section\\
    VV && Vertical-Vertical\\
    HH && Horizontal-Horizontal\\
    CFAR && Constant False Alarm Rate\\
    ROC && Region Of Convergence\\
    TPR && True Positive Rate\\
    FPR && False Positive Rate\\
    Log-Hough && Log-regularised Hough Transform\\
    \hline
    \end{tabular}}%
\end{center}

\section*{Acknowledgment}
We are grateful to the UK Satellite Applications Catapult for providing us the COSMO-SkyMed data sets employed in this study.
\bibliographystyle{IEEEtran}
\bibliography{GMC_journal}

\end{document}